\documentclass[%
 aip,
 amsmath,amssymb,
 reprint,%
]{revtex4-1}

\usepackage{graphicx}
\usepackage{dcolumn}
\usepackage{bm}

\usepackage[utf8]{inputenc}
\usepackage[T1]{fontenc}
\usepackage{mathptmx}
\usepackage{etoolbox}
\usepackage{bigints}
\usepackage[dvipsnames]{xcolor}

\makeatletter
\def\@email#1#2{%
 \endgroup
 \patchcmd{\titleblock@produce}
  {\frontmatter@RRAPformat}
  {\frontmatter@RRAPformat{\produce@RRAP{*#1\href{mailto:#2}{#2}}}\frontmatter@RRAPformat}
  {}{}
}%
\makeatother
\begin{document}

\preprint{AIP/123-QED}

\title[Effect of pressure on the CO$_{2}$ hydrate -- water interfacial energy]{Effect of pressure on the carbon dioxide hydrate -- water interfacial free energy along its dissociation line}

\author{Cristóbal Romero-Guzmán}
\affiliation{Laboratorio de Simulaci\'on Molecular y Qu\'imica Computacional, CIQSO-Centro de Investigaci\'on en Qu\'imica Sostenible and Departamento de Ciencias Integradas, Universidad de Huelva, 21006 Huelva Spain}

\author{Iv\'an M. Zer\'on}
\affiliation{Laboratorio de Simulaci\'on Molecular y Qu\'imica Computacional, CIQSO-Centro de Investigaci\'on en Qu\'imica Sostenible and Departamento de Ciencias Integradas, Universidad de Huelva, 21006 Huelva Spain}

\author{Jes\'us Algaba}
\affiliation{Laboratorio de Simulaci\'on Molecular y Qu\'imica Computacional, CIQSO-Centro de Investigaci\'on en Qu\'imica Sostenible and Departamento de Ciencias Integradas, Universidad de Huelva, 21006 Huelva Spain}

\author{Bruno Mendiboure}
\affiliation{Laboratoire des Fluides Complexes et Leurs R\'eservoirs, UMR5150, Universit\'e de Pau et des Pays de l’Adour, B.P. 1155, Pau Cdex 64014, France}

\author{Jos\'e Manuel M\'{\i}guez}
\affiliation{Laboratorio de Simulaci\'on Molecular y Qu\'imica Computacional, CIQSO-Centro de Investigaci\'on en Qu\'imica Sostenible and Departamento de Ciencias Integradas, Universidad de Huelva, 21006 Huelva Spain}

\author{Felipe J. Blas}
\affiliation{Laboratorio de Simulaci\'on Molecular y Qu\'imica Computacional, CIQSO-Centro de Investigaci\'on en Qu\'imica Sostenible and Departamento de Ciencias Integradas, Universidad de Huelva, 21006 Huelva Spain}
\email{felipe@uhu.es}

\begin{abstract}

We investigate the effect of pressure on the carbon dioxide (CO$_{2}$) hydrate--water interfacial free energy along its dissociation line using advanced computer simulation techniques. In previous works, we have determined the interfacial energy of the hydrate at $400 \,\text{bar}$ using the TIP4P/ice and TraPPE molecular models for water and CO$_{2}$, respectively, in combination with two different extensions of the Mold Integration technique [J. Chem. Phys. \textbf{141}, 134709 (2014)]. Results obtained from computer simulation, $29(2)$ and $30(2)\,\text{mJ/m}^{2}$, are found to be in excellent agreement with the only two measurements that exist in the literature, $28(6)\,\text{mJ/m}^{2}$ determined by Uchida \emph{et al.} [J. Phys. Chem. B \textbf{106}, 8202 (2002)] and $30(3)\,\text{mJ/m}^{2}$ by Anderson~\emph{et al.} [J. Phys. Chem. B \textbf{107}, 3507 (2002)]. Since the experiments  do not allow to obtain the variation of the interfacial energy along the dissociation line of the hydrate, we extend our previous studies to quantify the effect of pressure on the interfacial energy at different pressures. Our results suggest that there exists a correlation between the interfacial free energy values and the pressure, i.e., it decreases with the pressure between $100$ and $1000\,\text{bar}$. We expect that the combination of reliable molecular models and advanced simulation techniques could help to improve our knowledge of the thermodynamic parameters that control the interfacial free energy of hydrates from a molecular perspective.
\end{abstract}

\maketitle

%

\section{Introduction}

Natural gas hydrates are nonstoichiometric inclusion solid compounds in which guest molecules, such as methane (CH$_{4}$) or carbon dioxide (CO$_{2}$), are enclathrated in the voids left by a periodic network of water molecules (host). Small molecules, including CH$_{4}$ and CO$_{2}$, form type sI hydrates, while larger guests usually form type sII and type sH hydrate structures.~\cite{Sloan2008a} Applied research on hydrates has been motivated because these strategic materials are potential alternative sources of energy as CH$_{4}$ hydrate deposits,~\cite{Kvenvolden1988a,Koh2012a} and also are important from their global climate impact,~\cite{Sloan2008a,Manakov2003a,Makino2005a,Kvenvolden1988a,Koh2012a,Sloan2003a} CO$_{2}$ sequestration~\cite{Ohgaki1996a,Yang2014a,Ricaurte2014a} and gas storage~\cite{Kvamme2007a} and transportation.~\cite{Chihaia2005a,Peters2008a,English2009a}

From a fundamental point of view, understanding hydrate growth and nucleation is a challenging task.~\cite{Sloan2008a} Although some research groups have published significant results on hydrate nucleation from a molecular perspective, the molecular mechanisms that control it are still poorly understood.~\cite{Kashchiev2002a,Kashchiev2002b,Kashchiev2003a,Sarupria2011a,Walsh2011a,Sarupria2012a,Barnes2014b,Yuhara2015a} In this context, some of us have recently published some seminal papers to study the thermodynamic and kinetic parameters that control nucleation of methane and CO$_{2}$ hydrates from a molecular perspective.~\cite{Grabowska2022a,Grabowska2022b} One of these parameters is undeniably the interfacial free energy.~\cite{Aman2016a} In this work, we concentrate on the determination of the CO$_{2}$ hydrate--water interfacial free energy from computer simulation.

Experimental, theoretical, and molecular simulation techniques for determining fluid-fluid interfacial tensions are well established.~\cite{Adamson1997a,Hansen2013a,Allen2017a} However, there is a rather limited number of methods to obtain interfacial free energies of solid-fluid systems, with particular emphasis on the hydrate--water interfacial free energy.~\cite{Adamson1997a,Evans2005a} In the case of CO$_{2}$ hydrates, only two independent sets of experiments exist in the literature. Uchida \emph{et al.}~\cite{Uchida2002a} and Anderson \emph{et al.}~\cite{Anderson2003b} have obtained the interfacial free energy of the CO$_{2}$ hydrate from indirect measurements of the hydrate dissociation temperature in porous materials with varying pore size combined with the use of the Gibbs-Thomson equation.~\cite{Handa1992a,Uchida1999a,Uchida2002a,
Clennell1999a,Henry1999a,Anderson2003a} The analysis of the experimental data assumes that interfacial energy does not vary along the dissociation line when the pressure is changed, providing a single value of the interfacial energy.~\cite{Uchida2002a,Anderson2003b} In fact, the pressure dependence of the interfacial free energy is totally unknown experimentally and two important questions arise from these results after twenty years: (1) Does the interfacial free energy of a hydrate change along its dissociation line? And (2), if the answer to the former question is affirmative, how does the interfacial free energy vary along it?

Computer simulations of realistic models constitute an efficient alternative approach that can provide valuable information on hydrate--water interfacial energies from a microscopic point of view.~\cite{Sloan2008a} Very recently, we have determined the CO$_{2}$ hydrate--water interfacial energy by combining reliable molecular models for water and carbon dioxide and advanced molecular simulation techniques.~\cite{Algaba2022b,Zeron2022a} We have used two different but complementary extensions of the Mold Integration (MI) method proposed by Espinosa \emph{et al.}~\cite{Espinosa2014a} to determine the CO$_{2}$ hydrate--water interfacial free energy at $400 \,\text{bar}$. Particularly, we have used the TIP4P/Ice~\cite{Abascal2005b} and TraPPE~\cite{Potoff2001a} molecular models for water and CO$_{2}$, respectively, that are able to describe very accurately the CO$_{2}$ hydrate dissociation line in a wide range of pressures.~\cite{Miguez2015a} Calculations using the two extensions of the MI method also provide a good description of the interfacial free energy of the CO$_{2}$ hydrate at $400 \,\text{bar}$: $29(2) \,\text{mJ/m}^{2}$ and $30(2) \,\text{mJ/m}^{2}$ using the Mold Integration Host (MI--H)~\cite{Algaba2022b} and the Mold Integration Guest (MI--G),~\cite{Zeron2022a} respectively. These results are in excellent agreement with the experimental data of Uchida \emph{et al.}~\cite{Uchida2002a} and Anderson \emph{et al.}~\cite{Anderson2003b} for the interfacial energy, $26(8)$ and $30(3) \,\text{mJ/m}^{2}$.

In this work, we study the dependence of the CO$_{2}$ hydrate--water interfacial free energy, $\gamma_{hw}$ with the pressure along the three--phase coexistence line of the hydrate. We determine $\gamma_{hw}$ at different pressures using the MI--H technique and the same molecular models for water and CO$_{2}$ as in our previous works. It is important to remark that this election allows to describe very accurately the CO$_{2}$ hydrate dissociation line in a wide range of pressures, including the pressures considered in this study, and it provides an excellent description of the interfacial free energy of the CO$_{2}$ hydrate at $400\,\text{bar}$.

The organization of this paper is as follows: In Sec. II, we describe the methodology used in this work. Molecular simulation details are presented in Sec. III. The results obtained in this work are described in Sec. IV. Finally, conclusions are presented in Sec. V.


\section{Methodology}

The interfacial free energy of the CO$_{2}$ hydrate is calculated using the MI--H  methodology proposed by Algaba~\emph{et al.},~\cite{Algaba2022b} based on the original MI method proposed by Espinosa and collaborators.~\cite{Espinosa2014a} The method, which can be only applied at coexistence conditions, is based on the use of a mold of square-well interacting sites located in a lattice of one or several crystalline planes of the solid phase involved in the calculation of the interfacial energy.~\cite{Espinosa2014a,Espinosa2015a,Espinosa2016a} The mold can be switched on gradually to induce the formation of a solid slab in the fluid at coexistence conditions. This allows to evaluate the difference in free energy between the fluid and the fluid with the solid slab induced by the mold via thermodynamic integration. According to the method,~\cite{Espinosa2014a,Algaba2022b} the work needed to form the thin crystal slab of the solid phase in the fluid phase can be computed by integrating the average number of filled wells, $N_{fw}(\varepsilon)$, along the thermodynamic path linking both phases,

\begin{equation}
\Delta G^{hw}=N_{w}\epsilon_{m}-\int_{0}^{\varepsilon_{m}} \Big \langle N_{fw}(\varepsilon) \Big \rangle _{NP_{z}\mathcal{A}T}d\varepsilon \,
\label{eq:thermo-int}
\end{equation}

\noindent
where $N_{w}$ is the number of wells in the mold, $\varepsilon$ is the well depth, and $\varepsilon_{m}$ is the maximum well depth. The average number of the filled wells is obtained in the isothermal-isobaric ensemble to ensure that the system is always at the selected coexistence conditions. $\mathcal{A}$ represents the interfacial area between the hydrate and the aqueous solution of CO$_{2}$. In the isothermal-isobaric ensemble simulations used in this work, only fluctuations of the side perpendicular to the interface are allowed. Note that we have subtracted $-N_{w}\varepsilon_{m}$, the extra energy supplied to the system to induce the formation of the crystal slab. According to this, $\Delta G^{hw}$ represents the reversible work needed to create the slab and it allows the computation of the solid-fluid interfacial free energy in a straightforward fashion.

The success of the MI methodology depends on the election of several parameters that determine the final value of the interfacial free energy: the range of the square-well interaction sites of the mold, $r_{w}$, the number and distribution of the wells in the fluid phase, $N_{w}$, and the value of the maximum well depth, $\varepsilon_{m}$. In addition to that, the preparation of the initial simulation box is key since it determines if the system is at coexistence conditions when the integral in Eq.~\eqref{eq:thermo-int} is calculated. When $r_{w}$ is equal to the so-called optimal value,~\cite{Espinosa2014a} $r_{w}^{0}$,  $\Delta G^{hw}$ and the solid-fluid interfacial free energy,$\gamma_{hw}$, are related in a simple way,

\begin{equation}
\gamma_{hw}=\dfrac{\Delta G^{hw}}{2\mathcal{A}}
\label{interfacial-tension}
\end{equation}

\noindent
Factor 2 arises because two crystal-fluid interfaces appear when the solid slab is induced. In this work, we use the MI-H method to evaluate the CO$_{2}$ hydrate--water interfacial free energy. Since we use the MI--H method, the wells of the mold are located at the equilibrium crystallographic positions of the oxygen atoms of the water molecules of one layer of the sI structure of the hydrate.~\cite{Algaba2022b} The rest of the technical details of the methodology, the preparation of the initial simulation boxes, and how the number of  wells and layers, the maximum value of $\varepsilon$, and the optimal value $r_{w}^{0}$ are similar to those used in our seminal work.~\cite{Algaba2022b}

\section{Simulation details}

We use the GROMACS simulation package~\cite{VanDerSpoel2005a} to perform MD simulations via the $NP\mathcal{A}T$ or isothermal-isobaric ensemble~\cite{Allen2017a,Frenkel2002a} at $100$, $400$, and $1000 \,\text{bar}$ using the MI-H technique. Water and CO$_{2}$ molecules are modelled using the well-known TIP4P/Ice~\cite{Abascal2005b} and TraPPE~\cite{Potoff2001a} molecular models, respectively. We also use a modified Berthelot rule proposed by M\'{\i}guez~\emph{et al.}~\cite{Miguez2015a} Particularly, we use $\epsilon_{12}=\xi_{12}\,(\epsilon_{11}\,\epsilon_{22})^{1/2}$, with $\xi_{12}=1.13$ for the unlike Lennard-Jones interactions between CO$_{2}$ and water. The main reason for using this approach is to be consistent with our previous works~\cite{Algaba2022b,Zeron2022a} when comparing interfacial free energy values. Although this election overestimates the solubility of CO$_{2}$ in water, it provides an excellent estimation of the dissociation temperature of the CO$_{2}$ hydrate, $T_{3}$, along the whole dissociation line. The simulation box is a parallelepiped of volume $V=L_{x}\times L_{y}\times L_{z}$, with $L_{x}$, $L_{y}$, and $L_{z}$ the dimensions of the simulation box. $L_{x}$ and $L_{y}$ are kept constant and only $L_{z}$ is varied along the simulation. This ensures that the system is under the equilibrium normal pressure (perpendicular to the hydrate slab formed when the mold is switched on). To avoid stress in the slab of the CO$_{2}$ hydrate when the mold is switched on, we first simulate bulk solid phases using the anisotropic version of the $NPT$ ensemble to determine the equilibrium $L_{x}$ and $L_{y}$ dimensions of the simulation box. This ensures that the slabs simulated, at the equilibrium pressures, are non-stressed.~\cite{Noya2008a,Frenkel2013a,Espinosa2013a} 

To be consistent with our previous works,~\cite{Miguez2015a,Algaba2022b,Zeron2022a} the initial simulation boxes used in this work are prepared in an identical way. We use $736$ water molecules surrounded by two pure liquid slabs formed each of them by $128$ CO$_{2}$ molecules. Dimensions of the simulation boxes along the $x$ and $y$-axis parallel to the water-CO${_2}$ planar interface are chosen to be consistent with the size of the hydrate unit cell formed when the mold is switched on. We also use $N_{w}=48$ attractive sites, with $\varepsilon_{m}=8\,k_{B}T$, to induce the formation of the hydrate slab in the simulation box at the three pressures. An image of the initial configuration can be visualized in the supplementary material (multimedia files). The first frame of each movie corresponds to the initial configuration generated as explained previously.

Newton's equations are solved using a Verlet leapfrog algorithm~\cite{Cuendet2007a} with a time step of $0.002 \,\text{ps}$. All simulations are run at constant temperature and pressure using a Nos\'e-Hoover thermostat~\cite{Nose1984a} and a Parrinello-Rahman barostat.~\cite{Parrinello1981a} Simulations of the bulk solid phase to determine the equilibrium dimensions of the simulation box are run using the anisotropic version of the Parrinello-Rahman barostat. Relaxation times used in the thermostat and barostat are $2$ and $1 \,\text{ps}$, respectively. Long-range interactions due to electrostatic interactions are determined using the Particle Mesh Ewald technique.~\cite{Essmann1995a} We use a cutoff radius for both dispersive interactions and the real part of electrostatic interactions of $10 \,\text{\AA}$. Note that this cutoff distance is the same as that used in our previous works~\cite{Miguez2015a,Algaba2022b,Zeron2022a} and will allow to compare the results obtained with full consistency. No long range corrections were used in our study. LINCS algorithm is used in order to keep the molecular geometry. For an account of some additional simulation details, we recommend the revision of our previous work.~\cite{Algaba2022b}

\section{Results}

In order to find $r_{w}^0$ at different pressures, we run several simulations for different well radii and follow the evolution of $n_{h}$, the number of water molecules in the solid phase, when the mold is switched on at the beginning of the simulations. All of them start from a configuration of the CO$_{2}$ solution, at the appropriate concentration depending on the pressure, in contact with a reservoir of a pure CO$_{2}$ liquid phase as explained in the previous section. The number of water molecules forming the hydrate slab is calculated using rotational invariant local bond order parameters to distinguish between fluid-like and solid-like particles. We follow our previous works~\cite{Algaba2022b,Zeron2022a} and consider the combination of the $\overline{q}_{3}$ and $\overline{q}_{6}$ averaged order parameters proposed by Lechner and Dellago~\cite{Lechner2008a} to discriminate between fluid-like and solid-like water molecules during our simulations. All the details of the parameter used in this work to identify water molecules can be found in our previous papers.~\cite{Algaba2022b,Zeron2022a} We monitor $n_{h}=n_{h}(t)$ and identify the values of $r_{w}$ for which the curves do not exhibit an induction period and the hydrate grows, and those for which $n_{h}$ exhibits some induction period. The optimal value $r_{w}^{0}$, for each pressure, is enclosed between the two values for which the hydrate grows or does not grow.~\cite{Espinosa2014a,Algaba2022b,Zeron2022a}

\begin{figure}
\hspace*{-0.7cm}
\centering
\includegraphics[width=1.1\columnwidth]{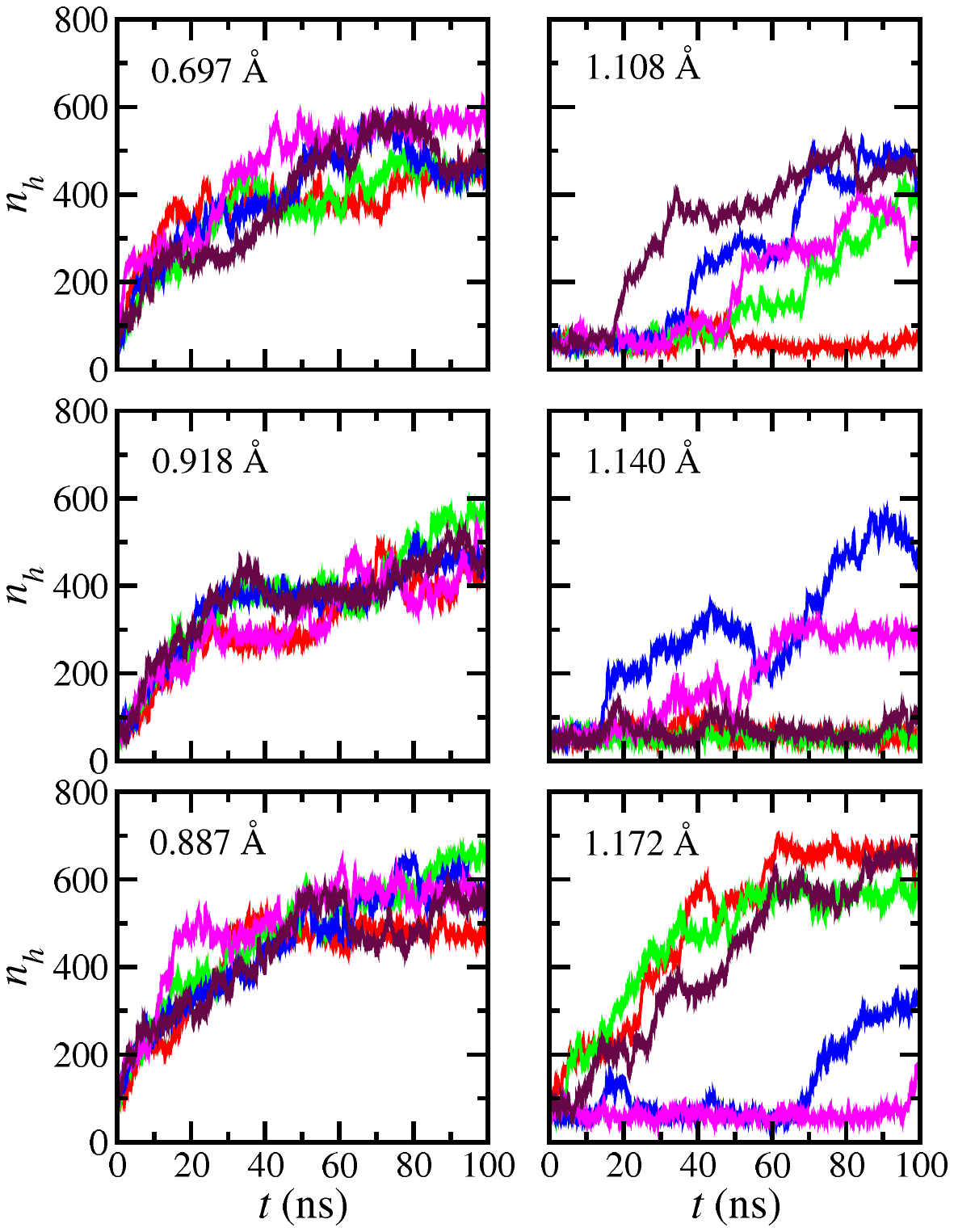}
\vspace*{-1.2cm}
\caption{Number of water molecules in the hydrate slab, $n_{h}$
 as a function of time for several trajectories and different well radius $r_{w}$ (as indicated in the legends). Simulations are performed using $\varepsilon = 8\,k_{B}T$ and at coexistence conditions: $100\,\text{bar}$ and $284\,\text{K}$ (upper panels), $400\,\text{bar}$ and $287\,\text{K}$ (middle panels), and $1000\,\text{bar}$ and $289\,\text{K}$ (bottom panels). Each color represents an independent trajectory generated using different seeds starting from the same fluid configuration.}
\label{figure1}
\end{figure}

Fig.~\ref{figure1} shows the evolution of $n_{h}$, as a function of time, of two representative values of $r_{w}$ for each pressure. As can be seen, for low values of $r_{w}$ the hydrate phase grows with no induction period in all trajectories. However, for high values of $r_{w}$ the hydrate does not grow. To determine the optimal values of $r_{w}$ it is necessary to get a more complete picture of the behavior of $n_{h}$, as a function of time, exploring several values of $r_{w}$. A complete representation of all the trajectories analyzed in this work (between 12 and 19 different $r_{w}$ values depending on the pressure considered) is presented in the supplementary material. 

It is not easy to determine precisely the optimal values of $r_{w}^0$ for hydrates. We follow here the criteria previously established by us in our previous works.~\cite{Algaba2022b,Zeron2022a} Particularly, at $1000 \,\text{bar}$, we find that $r_{w}=0.950 \,\text{\AA}$ is the highest $r_{w}$ value for which none of the trajectories shows an induction period. For $r_{w}=0.982 \,\text{\AA}$, at least one trajectory exhibits an induction period (see the supplementary material). Following the nomenclature used in our previous works,~\cite{Algaba2022b,Zeron2022a} we name these values as lower and upper bounds for $r_{w}^{0}$, $r_{w}^{(l)}=0.950\,\text{\AA}$ and $r_{w}^{(u)}=0.982\,\text{\AA}$.
According to this, the optimal value at $1000\,\text{bar}$ is $r_{w}^{0}=(r_{w}^{(u)}+r_{w}^{(l)})/2\approx 0.966 \,\text{\AA}$. As in our previous works, it is necessary to provide enough uncertainty to this value, which allows to give the appropriate confidence interval to $r_{w}^{0}$ and $\gamma_{hw}$. The uncertainty assigned to $r_{w}^{0}$
at this pressure is $0.08\,\text{\AA}$, and the final value at $1000\,\text{bar}$ is $r_{w}^{0}=0.97(8)\,\text{\AA}$. Following the same approach for the rest of pressures, we find that $r_{w}^{0}= 0.94(8)$ and $0.84(8)\,\text{\AA}$ for $400$ and $100\,\text{bar}$, respectively. See the complete series of $r_{w}$ values explored in this work in the supplementary material.

\begin{figure}
\hspace*{-0.6cm}
\centering
\includegraphics[width=1.1\columnwidth]{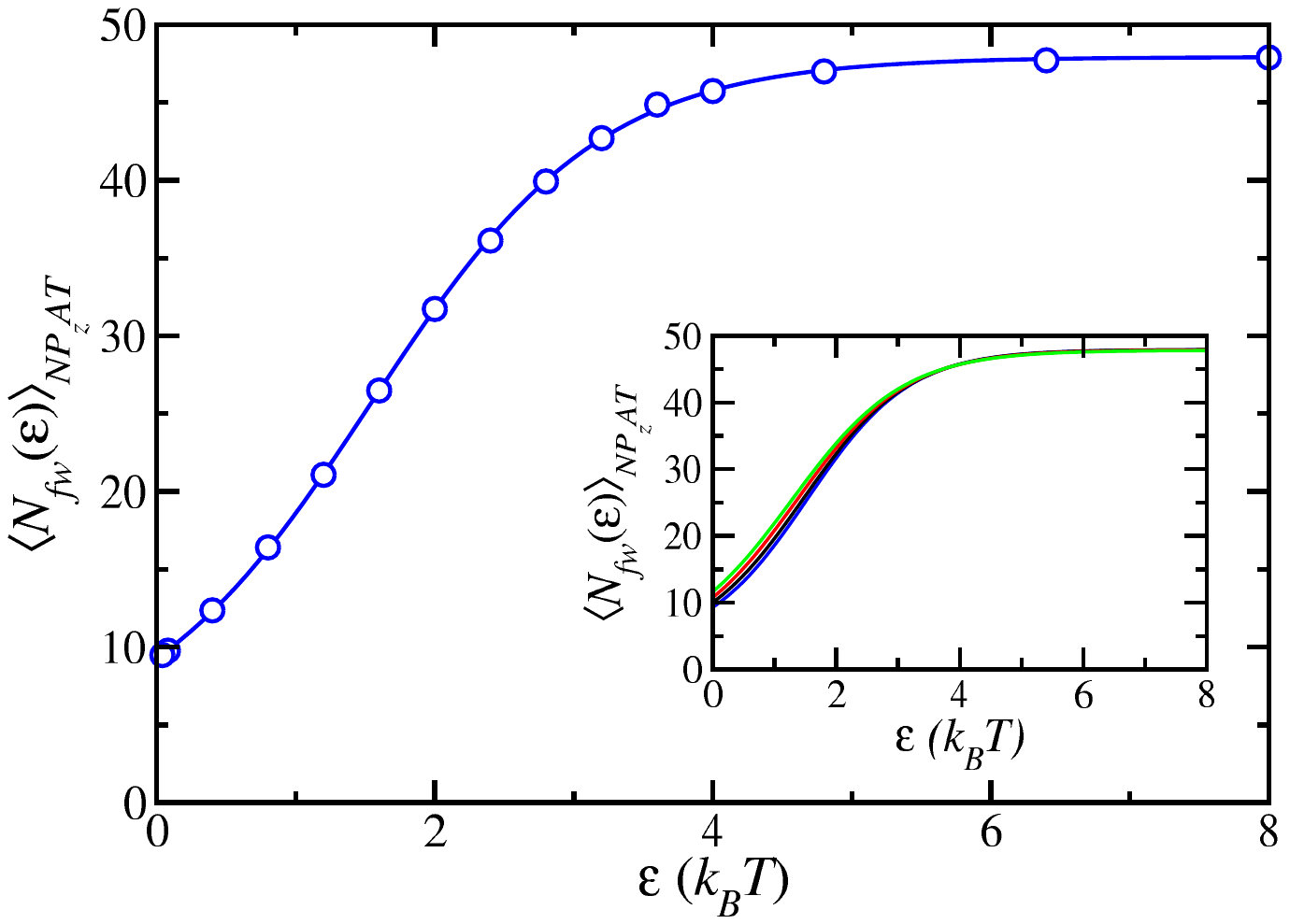}
\vspace*{-0.8cm}
\caption{The average number of filled wells, 
$\big \langle N_{fw}(\varepsilon)\big \rangle_{NP_{z}\mathcal{A}T}$, as a function of the well depth $\varepsilon$, at $100\,\text{bar}$ for the principal plane of the CO$_{2}$ hydrate. The radius of the mold used is $r_{w}=1.140\,\text{\AA}$. The circles correspond to the values obtained from $NP_{z}\mathcal{A}T$ simulations of $40\,\text{ns}$ and the blue curve represents its corresponding fit. The inset represents the fits of the average number of filled wells using well radii higher than the optimal value, $r_{w}=1.140$ (blue), $1.172$ (black), $1.203$ (red), and $1.235\,\text{\AA}$ (green).}
\label{figure2}
\end{figure}

According to the previous paragraphs, the induction period identified should be related with the absence of a crystal structure around the mold. According to the original methodology, this implies that the system is in an induction period prior to the overcoming of a free energy barrier. However, another interpretation is possible. The absence of crystal structure could be also interpreted as a stuck crystal growth due to the low step-like kinetics with which the hydrate apparently grows. Unfortunately, it is very difficult to distinguish between induction period and stuck growth at the beginning of the simulations in which there is absence of crystal structure. A detailed analysis of this alternative way to evaluate the optimal value of $r_{w}$ is presented in the supplementary material.

Once the optimal values $r_{w}^{0}$ for each pressure have been determined, it is possible to calculate $\gamma_{hw}$ for several values $r_{w}>r_{w}^{0}$. Particularly, we use four different values of $r_{w}$ at each pressure to calculate the integral in Eq.~\eqref{eq:thermo-int}. This is done by determining the average number of filled wells with water molecules for a given number of $\varepsilon$ values. Fig.~\ref{figure2} shows $\big \langle N_{fw}(\varepsilon) \big \rangle _{NP_{z}\mathcal{A}T}$, as a function of $\varepsilon$, at $100\,\text{bar}$ using a value $r_{w}=1.140\,\text{\AA}$. We have also considered other values of the potential range of the wells, $r_{w}=1.172$, $1.203$, and $1.235\,\text{\AA}$, shown in the inset of Fig.~\ref{figure2}. Note that different values for $r_{w}$ give slightly different filling curves. All these values are larger than the optimal value to ensure that the states visited by the system during the simulations correspond to liquid states. With this election, the integration of Eq.~\eqref{eq:thermo-int} is always reversible. Note that $\big \langle N_{fw}(\varepsilon) \big \rangle _{NP_{z}\mathcal{A}T}$, as a function of time, behaves smoothly and it reaches the expected plateau value ($N_{fw}\rightarrow N_{w}$) when $\varepsilon\rightarrow\varepsilon_{m}$. Finally, we have also checked that all the wells of the mold are fully occupied with only one water molecule for all the values of $\varepsilon$ used for the integration. The rest of the filling curves corresponding to our calculations at $400$ and $1000\,\text{bar}$ are presented in the supplementary material.

\begin{figure}
\hspace*{-0.4cm}
\centering
\includegraphics[width=1.1\columnwidth]{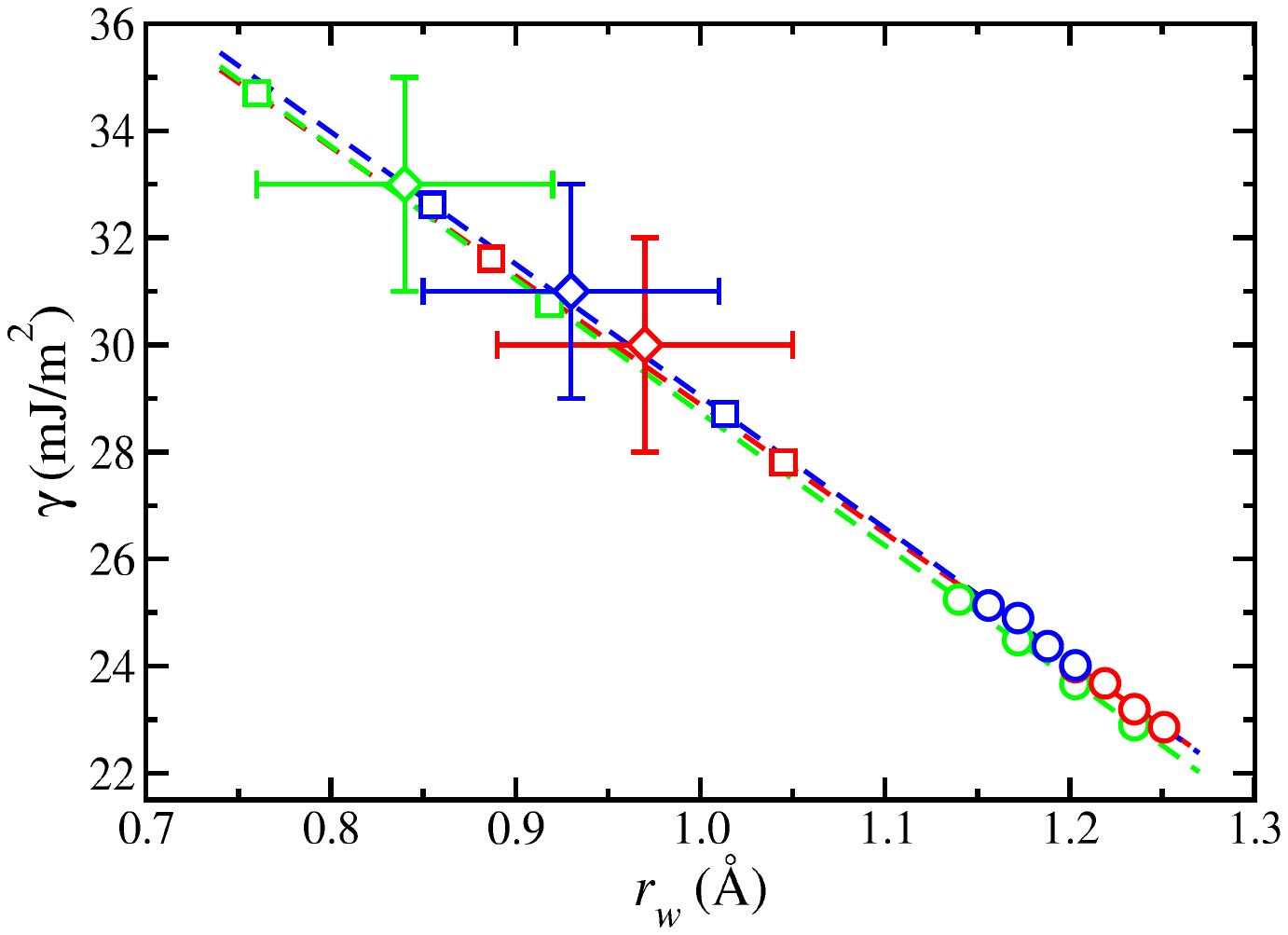}
\vspace*{-0.8cm}
\caption{CO$_{2}$ hydrate--water interfacial free energy as a function of the potential well radius of the mold as obtained from the MI--H methodology at $100$ (open green circles), $400$ (open blue circles), and $1000\,\text{bar}$ (open red circles). The dashed lines represent linear fits of each date at the corresponding pressure, the open squares the interfacial tensions evaluated at the lower and upper bounds of $r_{w}^{0}$, $r_{w}^{(l)}$ and $r_{w}^{(u)}$, at each pressure, and the open diamonds the extrapolation of the linear fit to the optimal well radius at the corresponding pressure.}
\label{figure3}
\end{figure}

The CO$_{2}$ hydrate--water interfacial free energy values, for each pressure and different choices of $r_{w}$, can be readily obtained by integrating the corresponding $\big \langle N_{fw}(\varepsilon) \big \rangle _{NP_{z}\mathcal{A}T}$ functions with respect to $\varepsilon$ according to Eq.~\eqref{eq:thermo-int} and using Eq.~\eqref{interfacial-tension}. The results obtained are shown in Fig.~\ref{figure3}. As can be seen, the interfacial free energy values for each pressure (open circles) follow a linear behavior with $r_{w}$, in agreement with previous results.~\cite{Espinosa2014a,Espinosa2016a,Algaba2022b,Zeron2022a} We also perform linear fits of the $\gamma_{hw}$ values (for $r_{w}<r_{w}^{0}$), at each pressure, to obtain the value of the interfacial energy evaluated at each optimal value $r_{w}^{0}$. Following our previous works, we have estimated the uncertainties of $\gamma_{hw}$ at each pressure from the uncertainties of $r_{w}^{0}$. Particularly, the error associated with each value of $\gamma_{hw}$ is half of the difference between the values of the interfacial energy evaluated at the lower and upper bounds of $r_{w}$ around each optimal value (open squares). 

The interfacial free energy value that we obtain in this work at $400\,\text{bar}$, using $N_{w}=48$ wells for the mold, $31(2)\,\text{mJ/m}^{2}$, is fully consistent with those obtained in our previous works,~\cite{Algaba2022b,Zeron2022a} $29(2)\,\text{mJ/m}^{2}$ and $30(2)\,\text{mJ/m}^{2}$. According to this, we obtain the same value of the CO$_{2}$ hydrate--water interfacial energy within the error bars. This result is especially interesting since we have previously used two different versions of the method and different number of wells, the MI-H method with $N_{w}=56$ sites for water molecules and the MI-G technique with $N_{w}=16$ sites for the CO$_{2}$ molecules. The interfacial free energy values at $100$ and $1000\,\text{bar}$ are $33(2)$ and $30(2)\,\text{mJ/m}^{2}$, respectively. To the best of our knowledge, this is the first time the interfacial free energy of the CO$_{2}$ hydrate-water interface is determined from computer simulation along its dissociation line. 

\begin{figure}
\hspace*{-0.4cm}
\centering
\includegraphics[width=1.1\columnwidth]{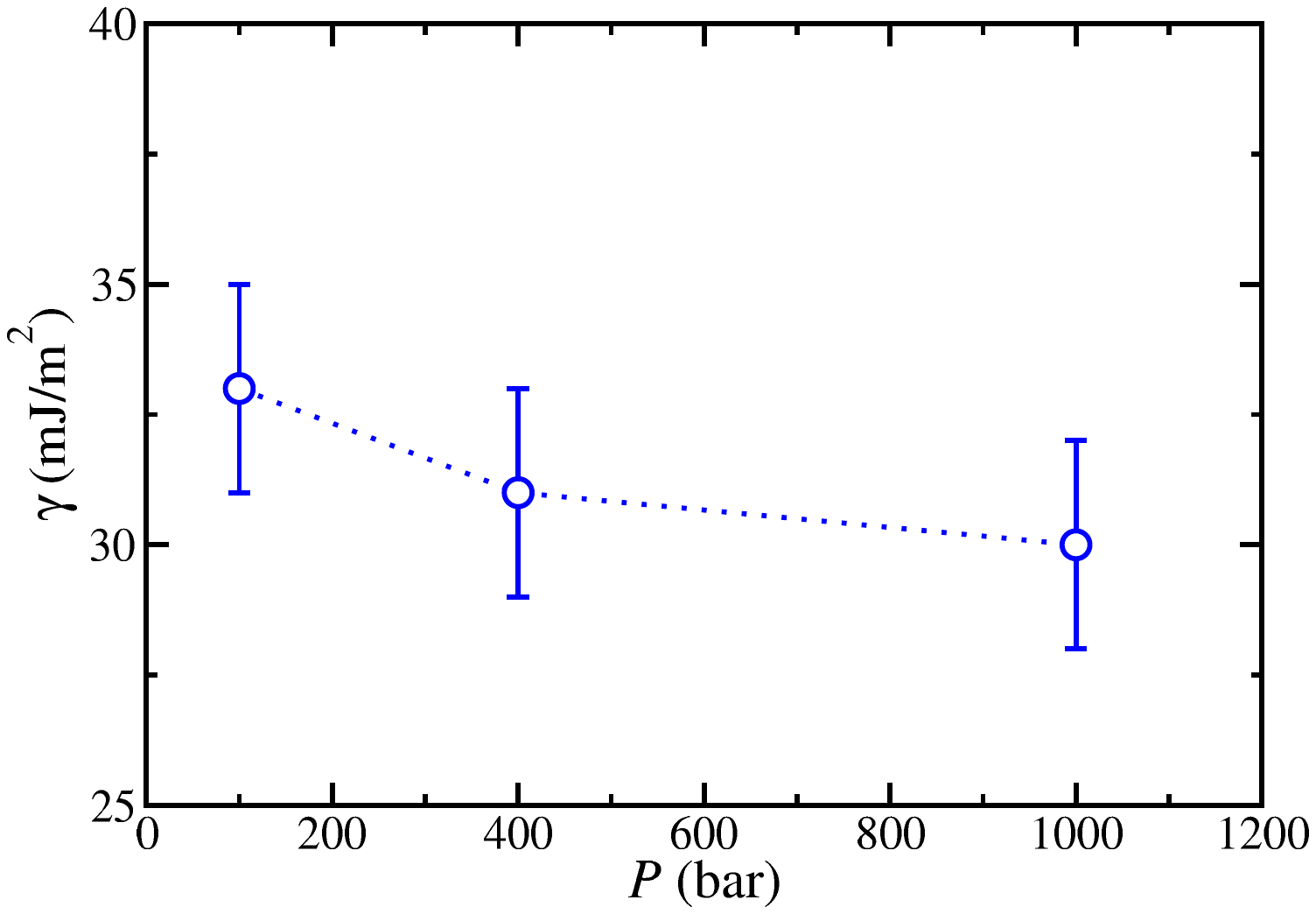}
\vspace*{-0.4cm}
\caption{CO$_{2}$ hydrate--water interfacial free energy as a function of pressure along its dissociation line as obtained from the MI--H methodology. The dotted curve is a guide to the eye.}
\label{figure4}
\end{figure}

It is also interesting to compare the interfacial energy values for varying pressure with the experimental data of Uchida~\emph{et al.}~\cite{Uchida2002a} and Anderson~\emph{et al.}~\cite{Anderson2003b} obtained independently several years ago. Our results for $\gamma_{hw}$, at the three pressures, agree very well with the value proposed by Uchida and collaborators, $28(6)\,\text{mJ/m}^{2}$, and also with those measured by Anderson and coworkers, $30(3)\,\text{mJ/m}^{2}$. We should notice that error bars are large, not only for the experimental values, $6$ and $3\,\text{mJ/m}^{2}$, but also for the simulation data obtained in this work, $2\,\text{mJ/m}^{2}$. A first analysis of the results seems to indicate that the interfacial free energy is constant in a wide range of pressures, between $100$ and $1000\,\text{bar}$. This is in agreement with the data provided by Uchida~\emph{et al.}~\cite{Uchida2002a} and Anderson~\emph{et al.}~\cite{Anderson2003b} if we take into account that all the results are within the error bars. However, is the interfacial free energy of a hydrate constant along its dissociation line?

The experimental data taken from the literature were published without specifying the pressures at which $\gamma_{hw}$ values were obtained.~\cite{Uchida2002a,Anderson2003b} More precisely, the authors estimated the interfacial free energy of the CO$_{2}$ hydrate from pressure-temperature dissociation experimental data measured in porous materials combined with the phenomenological Gibbs-Thomson equation.~\cite{Handa1992a,
Clennell1999a,Henry1999a} They assumed, among other approximations, that the interfacial free energy is constant and it only depends on the mean diameter of the pores. According to this, the hydrate interfacial free energy does not vary with the pressure. In other words, as it is mentioned in the Introduction, the dependence of $\gamma_{hw}$ with pressure is not known experimentally.

However, the technique used in this work allows to obtain $\gamma_{hw}$ at different pressures along the hydrate dissociation line at coexistence conditions.
We have represented the interfacial free energy as a function of pressure in Fig.~\ref{figure4}. As can be seen, our results suggest that there is a weak correlation between the interfacial energy values and the pressure, i.e., $\gamma_{hw}$ decreases with pressure along the dissociation line.  To the best of our knowledge, this behavior has not been previously reported in the literature and it is a surprising result. The dependence of the solid-liquid interfacial free energy with pressure has been studied in the literature by two groups in two different systems. On one hand, Laird \emph{et al.} have determined the solid-liquid interfacial free energy along the coexistence line of the Lennard-Jones (LJ) system using Gibbs-Cahn integration.~\cite{Laird2009a}  On the other hand, Espinosa~\emph{et al.} have calculated the water--Ih ice interfacial free energy at two different pressures, $1$ and $2000\,\text{bar}$, on the melting line of Ih ice, using the original MI method.~\cite{Espinosa2016a,Espinosa2016d} Although the pressure--temperature slope of the corresponding coexistence lines have different sign (in the LJ system the slope is positive and in the water case is negative), in both cases the interfacial free energy increases with pressure. This behavior is contrary to that observed in the case of the CO$_{2}$ hydrate. Note that the dissociation line of the CO$_{2}$ hydrate exhibits a positive slope in the range of pressures studied in this work.

It is important to recall here that the optimal values of $r_{w}$, at different pressures, can be calculated using two different criteria (see supplementary material for further details). Both criteria show the same qualitative behavior: there is a weak correlation between the interfacial energy values and the pressure, i.e., $\gamma_{hw}$ slightly decreases with pressure along the dissociation line. Finally, it is interesting to mention that results obtained from the original and overall behavior criteria predict the same results within the error bars. However, further work is needed to find a more exact method to reduce the ambiguity with which the optimal radius is chosen in the Mold Integration method. This will be the subject of a future work.

It would be interesting to explain this behavior from a molecular perspective. In the case of the water--Ih ice interface, Espinosa and coworkers~\cite{Espinosa2016d} proposed that the interfacial energy increase could be due to the breakage of hydrogen bonds in the liquid phase. It seems this is not the case for the CO$_{2}$ hydrate. However, this is not a definitive conclusion. The error bars associated to the experimental measurements are large, as well as those corresponding to the simulation data. A more detailed study would provide additional information on the behavior of the CO$_{2}$ hydrate interfacial free energy as a function of the pressure along its dissociation line.

\section{Conclusions}

We have analyzed the effect of pressure on the CO$_{2}$~hydrate--water interfacial free energy along its dissociation line from MD computer simulations. We have used the TraPPE force field to model CO$_{2}$ molecules and the TIP4P/Ice model to describe H$_{2}$O molecules. This election allows to describe very accurately the CO$_{2}$ hydrate dissociation line in a wide range of pressures, including the pressures considered in this study. 

In two previous papers, we have used the same models and two extensions of the original MI methodology to account for the interfacial energy of the CO$_{2}$ hydrate at $400\,\text{bar}$. These techniques, which allow to calculate the free energy at coexistence conditions, are based on the definition of the interfacial free energy and well-established tools from Thermodynamics and Statistical Mechanics. In this work, we use one of the extensions, the co-called MI--H technique, to determine the interfacial free energy at three different pressures along the dissociation line, $100$, $400$, and $1000\,\text{bar}$.

The interfacial free energy value obtained at $400\,\text{bar}$, $31(2)\,\text{mJ/m}^{2}$, is fully consistent with those obtained previously using two independent versions of the method, $29(2)$ and $30(2)\,\text{mJ/m}^{2}$. We also obtain similar values of the interfacial energy at $100$ and $1000\,\text{bar}$, $33(2)$ and $30(2)\,\text{mJ/m}^{2}$, respectively. All the results are consistent and agree very well with the experimental values taken from the literature, $28(6)$ and $30(3)\,\text{mJ/m}^{2}$.

The results obtained seem to indicate in a first analysis that the interfacial free energy is constant in a wide range of pressures, in agreement with the results provided by experiments. Unfortunately, the experimental technique assumes that the interfacial energies are independent of the pressure. Contrarily, the MI--H and MI--G methods allow to determine the interfacial free energy at different thermodynamic conditions at coexistence. A more careful analysis of the results suggests that there is a weak correlation between the interfacial free energy values and the pressure, i.e., $\gamma_{hw}$ decreases with the pressure. This is not a definitive conclusion since the error bars associated with the experimental measurements are large, as well as those corresponding to the simulation data, and our study has been performed in a limited range of pressures, between $100$ and $1000\,\text{bar}$. Although a more detailed analysis is needed to confirm the main result of this study, this work corroborates that the combination of the TIP4P/ice and TraPPE models and the MI methodology is a powerful and valuable approach for calculating interfacial free energies of CO$_{2}$ hydrates.

\section*{Supplementary material}
See the supplementary material for the time evolution of the number of water molecules in the CO$_{2}$ hydrate slab for the different well radii and pressures considered, the filling curves calculated at $400$ and $1000\,\text{bar}$, and the multimedia results (movies) obtained from MD computer simulations showing crystallization and no crystallization starting from an equilibrated liquid–liquid configuration of the CO$_{2}$ + H$_{2}$O binary mixture.

\section*{Acknowledgements}
We thank Carlos Vega for helpful discussions. This work was financed by Ministerio de Ciencia e Innovaci\'on (Grant No.~PID2021-125081NB-I00), Junta de Andalucía (P20-00363), and Universidad de Huelva (P.O. FEDER UHU-1255522 and FEDER-UHU-202034), all four co-financed by EU FEDER funds. We also acknowledge the Centro de Supercomputaci\'on de Galicia (CESGA, Santiago de Compostela, Spain) for providing access to computing facilities.

\section*{Author declarations}

\noindent
\textbf{Conflict of interests}

The authors declare no conflicts to disclose.

\section*{Author contributions}

\noindent
\textbf{Crist\'obal Romero-Guzm\'an:} Visualization (lead); Methodology (equal); Investigation (lead); Writing – review \& editing (equal).
\textbf{Iván M. Zerón:} Methodology (equal); Writing – review \& editing (equal).
\textbf{José Manuel Míguez:} Validation (lead); Methodology (equal); Writing – review \& editing (equal).
\textbf{Bruno Mendiboure:} Formal analysis (lead); Methodology (equal); Writing – review \& editing (equal).
\textbf{Jesús Algaba:} Methodology (equal); Writing – review \& editing (equal).
\textbf{Felipe J. Blas:} Conceptualization (lead); Funding acquisition (lead); Methodology (equal); Writing – original draft (lead); Writing – review \& editing (equal).

\section*{Data availability}

The data that support the findings of this study are available within the article and its supplementary material.

\bibliography{bibfjblas}

\end{document}



\begin{center}
\huge Supplementary Material
\end{center}

\title{Effect of pressure on the carbon dioxide hydrate -- water interfacial free energy along its dissociation line}

\author{Cristóbal Romero-Guzmán}
\affiliation{Laboratorio de Simulaci\'on Molecular y Qu\'imica Computacional, CIQSO-Centro de Investigaci\'on en Qu\'imica Sostenible and Departamento de Ciencias Integradas, Universidad de Huelva, 21006 Huelva Spain}

\author{Iv\'an M. Zer\'on}
\affiliation{Laboratorio de Simulaci\'on Molecular y Qu\'imica Computacional, CIQSO-Centro de Investigaci\'on en Qu\'imica Sostenible and Departamento de Ciencias Integradas, Universidad de Huelva, 21006 Huelva Spain}

\author{Jes\'us Algaba}
\affiliation{Laboratorio de Simulaci\'on Molecular y Qu\'imica Computacional, CIQSO-Centro de Investigaci\'on en Qu\'imica Sostenible and Departamento de Ciencias Integradas, Universidad de Huelva, 21006 Huelva Spain}

\author{Bruno Mendiboure}
\affiliation{Laboratoire des Fluides Complexes et Leurs R\'eservoirs, UMR5150, Universit\'e de Pau et des Pays de l’Adour, B.P. 1155, Pau Cdex 64014, France}

\author{Jos\'e Manuel M\'{\i}guez}
\affiliation{Laboratorio de Simulaci\'on Molecular y Qu\'imica Computacional, CIQSO-Centro de Investigaci\'on en Qu\'imica Sostenible and Departamento de Ciencias Integradas, Universidad de Huelva, 21006 Huelva Spain}

\author{Felipe J. Blas}
\affiliation{Laboratorio de Simulaci\'on Molecular y Qu\'imica Computacional, CIQSO-Centro de Investigaci\'on en Qu\'imica Sostenible and Departamento de Ciencias Integradas, Universidad de Huelva, 21006 Huelva Spain}
\email{felipe@uhu.es}

\maketitle

%

\section{Time evolution of number of water molecules in hydrate}

According to the main article, we have determined the time evolution of $n_{h}$, the number of water molecules in the CO$_{2}$ hydrate slab, at three different pressures along its dissociation line. Figs.~1, 2, and 3 show $n_{h}$, as a function of time, at $100\,\text{bar}$ for $r_{w}$ ranging from $0.697$ to $1.235\,\text{\AA}$. Fig.~4 shows $n_{h}$, as a function of time, at $400\,\text{bar}$ for $r_{w}$ ranging from $0.918$ to $1.203\,\text{\AA}$. And finally, Figs.~5 and 6 show $n_{h}$, as a function of time, at $1000\,\text{bar}$ for $r_{w}$ ranging from $0.887$ to $1.267\,\text{\AA}$.

In all cases, it is possible to identify three different scenarios, as defined in our previous works~\cite{Algaba2022b,Zeron2022a}: (a) Scenario I, in which none of the trajectories exhibits induction period and system always crystallizes if simulation runs are sufficiently long (no free energy barrier between the fluid and solid phases exists); (b) Scenario II,  in which at least one trajectory shows an induction period, i.e., $n_{h}$ does not grow, on average, from $t=0\,\text{ns}$; and (c) Scenario III, in which none of them exhibit, on average, monotonic increasing of $n_{h}$ with time due to the existence of a free energy barrier between the fluid and solid that does not allow the system to crystallize. As can be seen, there exists a minimum value of $r_{w}$ for which at least one of the trajectories shows an induction period at different pressures: (a) $100\,\text{bar}$ for $r_{w}\ge 0.855\,\text{\AA}$ (see Fig.~1); (b) $400\,\text{bar}$ for $r_{w}\ge 0.950\,\text{\AA}$ (see Fig.~4); and $1000\,\text{bar}$ for $r_{w}\ge 0.982\,\text{\AA}$ (see Fig.~5).

\begin{figure}
\centering
\includegraphics[angle=0,width=1.00\columnwidth]{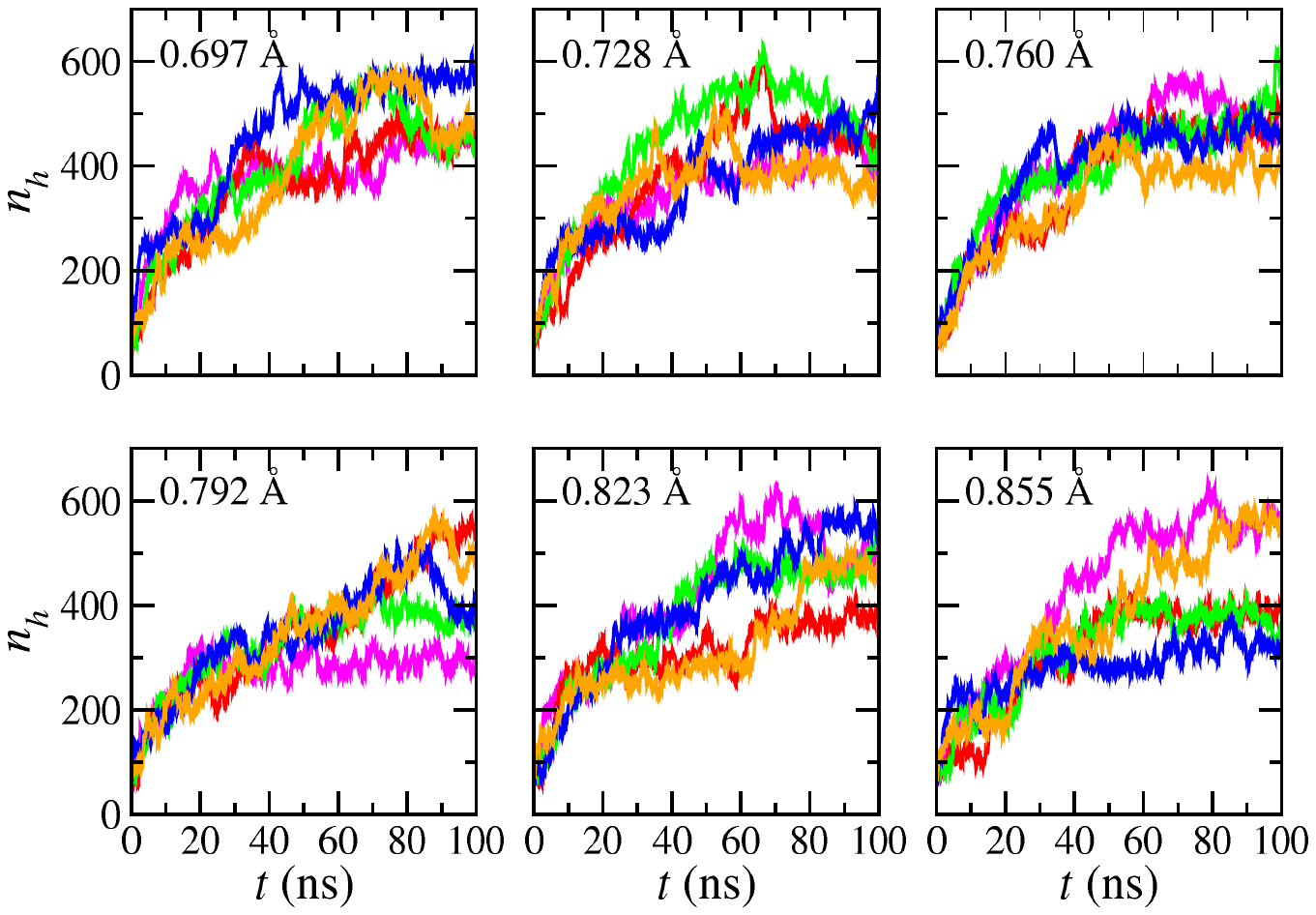}
\caption{\small Number of water molecules in the crystal slab, $n_{h}$, as function of time for several trajectories and different well radius, $r_{w}$ (as indicated in the legends). All simulations are performed at coexistence conditions ($100\,\text{bar}$ and $284\,\text{K}$). In all cases, $\varepsilon=8\,\text{k}_{B}T$. Each color represents an independent trajectory generated using different seeds starting from the same fluid configuration.
}
\label{s1}
\end{figure}

\clearpage

\begin{figure}
\centering
\includegraphics[angle=0,width=1.00\columnwidth]{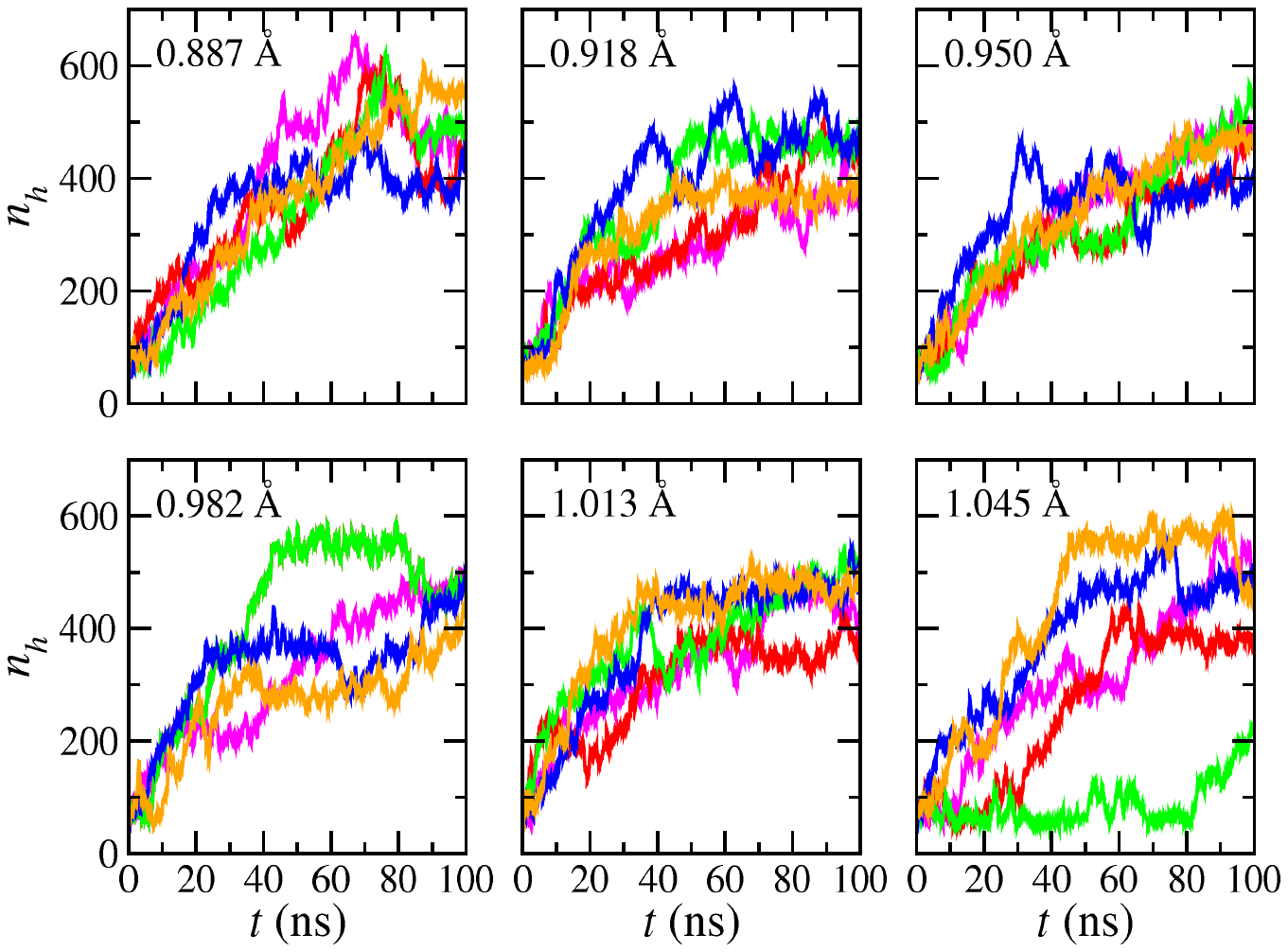}
\caption{\small Number of water molecules in the crystal slab, $n_{h}$, as function of time for several trajectories and different well radius, $r_{w}$ (as indicated in the legends). All simulations are performed at coexistence conditions ($100\,\text{bar}$ and $284\,\text{K}$). In all cases, $\varepsilon=8\,\text{k}_{B}T$. Each color represents an independent trajectory generated using different seeds starting from the same fluid configuration.
}
\label{s2}
\end{figure}

\clearpage

\begin{figure}
\centering
\includegraphics[angle=0,width=1.0\columnwidth]{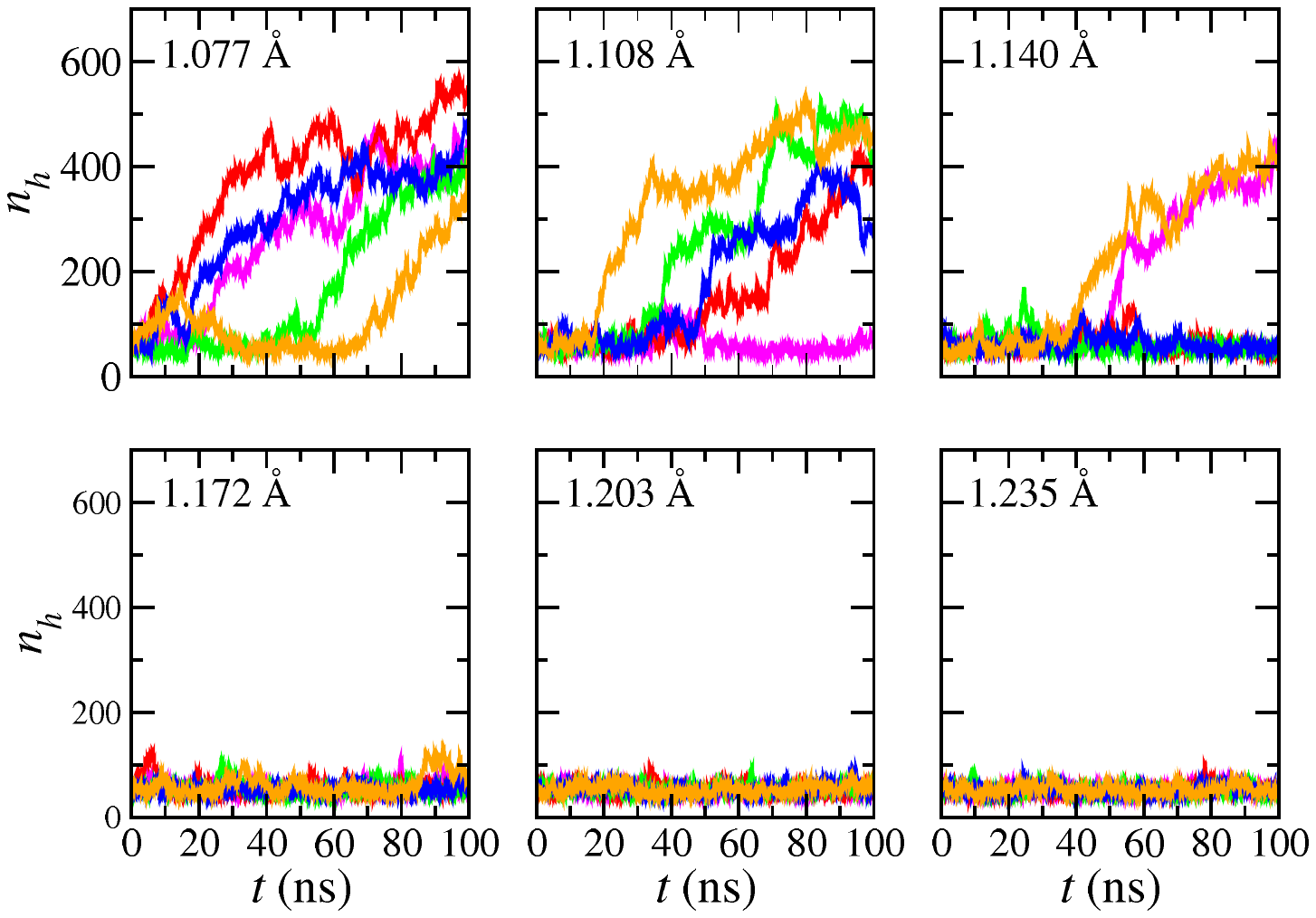}
\caption{\small Number of water molecules in the crystal slab, $n_{h}$, as function of time for several trajectories and different well radius, $r_{w}$ (as indicated in the legends). All simulations are performed at coexistence conditions ($100\,\text{bar}$ and $284\,\text{K}$). In all cases, $\varepsilon=8\,\text{k}_{B}T$. Each color represents an independent trajectory generated using different seeds starting from the same fluid configuration.
}
\label{s3}
\end{figure}

\clearpage

\begin{figure}
\centering
\includegraphics[angle=0,width=1.0\columnwidth]{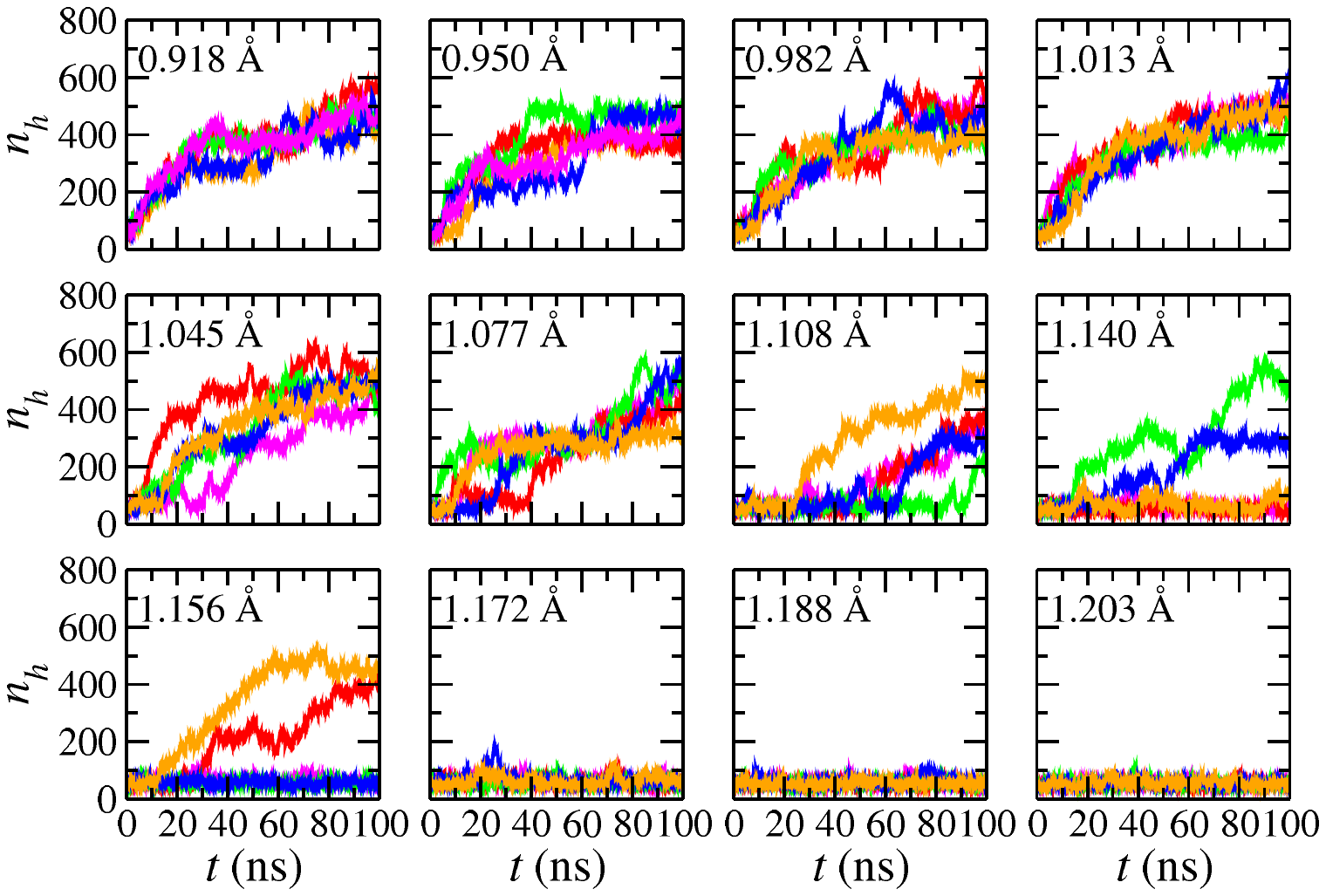}
\caption{\small Number of water molecules in the crystal slab, $n_{h}$, as function of time for several trajectories and different well radius, $r_{w}$ (as indicated in the legends). All simulations are performed at coexistence conditions ($400\,\text{bar}$ and $287\,\text{K}$). In all cases, $\varepsilon=8\,\text{k}_{B}T$. Each color represents an independent trajectory generated using different seeds starting from the same fluid configuration.
}
\label{s4}
\end{figure}

\clearpage

\begin{figure}
\centering
\includegraphics[angle=0,width=1.00\columnwidth]{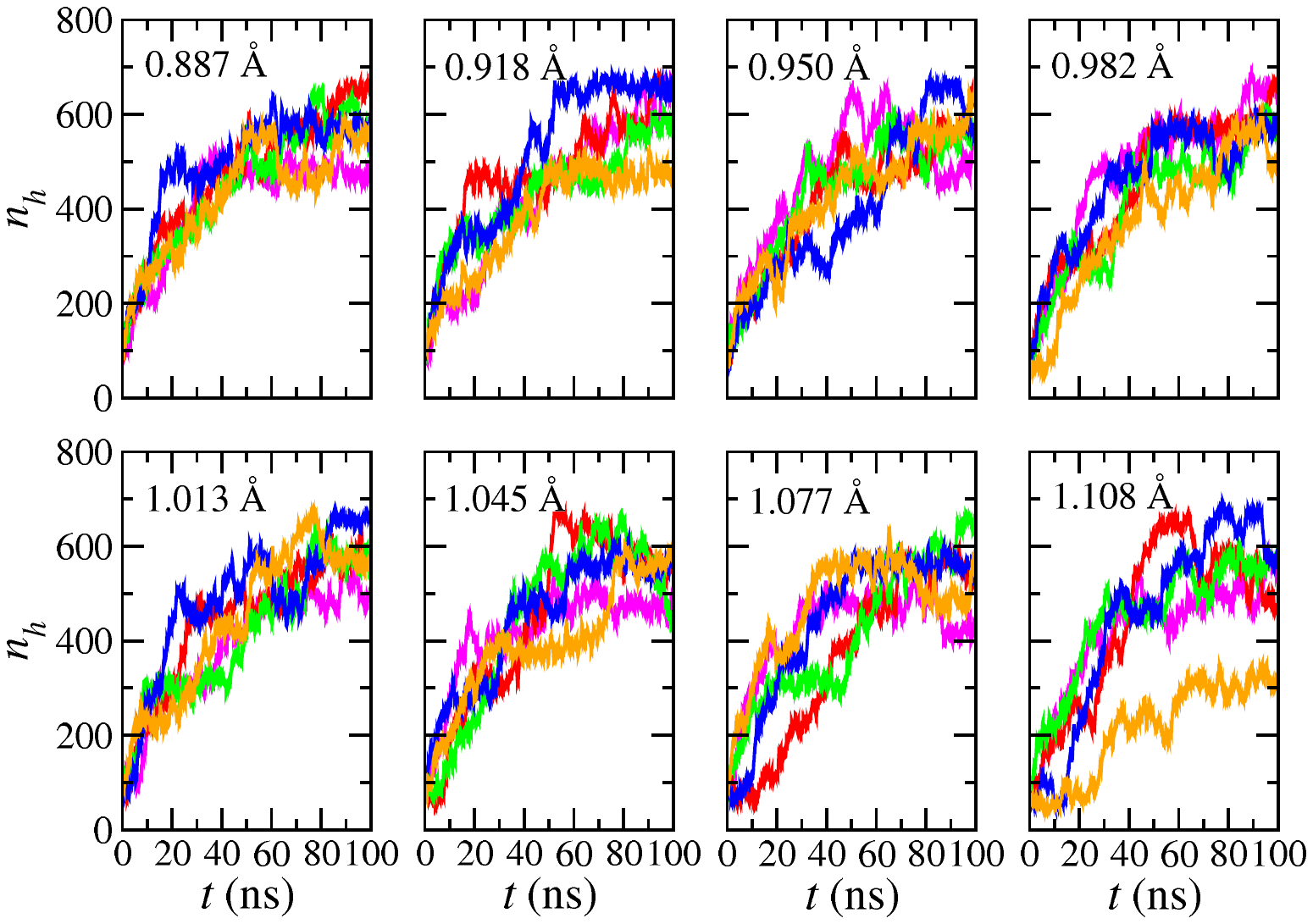}
\caption{\small Number of water molecules in the crystal slab, $n_{h}$, as function of time for several trajectories and different well radius, $r_{w}$ (as indicated in the legends). All simulations are performed at coexistence conditions ($1000\,\text{bar}$ and $289\,\text{K}$). In all cases, $\varepsilon=8\,\text{k}_{B}T$. Each color represents an independent trajectory generated using different seeds starting from the same fluid configuration.
}
\label{s5}
\end{figure}

\clearpage

\begin{figure}
\centering
\includegraphics[angle=0,width=1.00\columnwidth]{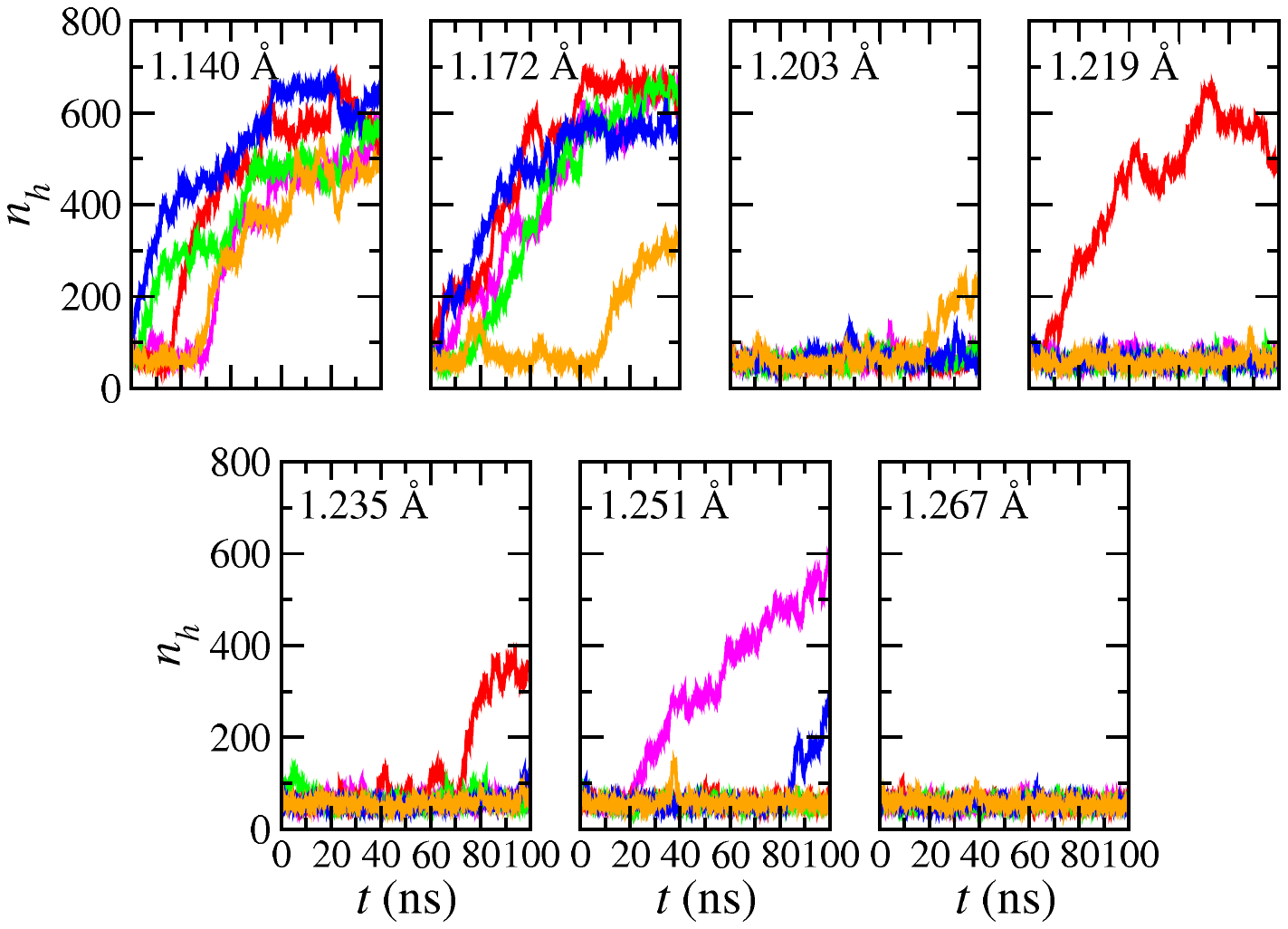}
\caption{\small Number of water molecules in the crystal slab, $n_{h}$, as function of time for several trajectories and different well radius, $r_{w}$ (as indicated in the legends). All simulations are performed at coexistence conditions ($1000\,\text{bar}$ and $289\,\text{K}$). In all cases, $\varepsilon=8\,\text{k}_{B}T$. Each color represents an independent trajectory generated using different seeds starting from the same fluid configuration.
}
\label{s6}
\end{figure}

\clearpage

\section{Election of the optimal value}

\subsection{Original criterion}

According to the original works of Espinosa \emph{et al.},~\cite{Espinosa2014a,Espinosa2016a} the optimal value  $r_{w}^{0}$ is located between the highest value $r_{w}$ for which none of the trajectories shows an induction period and the lowest value for which at least one trajectory exhibits an induction period. We have followed this approach in the main article to be consistent with the original method and with our previous works. The different $r_{w}^{0}$ values obtained following the original criterion are presented in Table~\ref{table}.

\begin{table}[h]
\caption{\label{table}%
Optimal values, $r_{w}^{0}$, as obtained using the original and overall behavior criteria at different pressures. All simulations are performed using $\epsilon=8\,k_{B}T$ and at coexistence conditions. See further details in the main text.
}
\begin{ruledtabular}
\begin{tabular}{cccc}
\textrm{$P$ (bar)} & \textrm{$r_{w}^{0}$ (\AA)\footnote{Original criterion.}}&
\textrm{$r_{w}^{0}$} (\AA)\footnote{Overall behavior criterion using the data from regions I and II.}
& \textrm{$r_{w}^{0}$} (\AA)\footnote{Overall behavior criterion using only the data from region II.}\\
\colrule
100 &  0.84(8) & 0.87(8) & 1.00(2)\\
400 &  0.94(8) & 0.97(5) & 1.03(2)\\
1000 & 0.97(8) & 0.99(4) & 1.04(3)\\
\end{tabular}
\end{ruledtabular}
\end{table}

Our election is motivated by reasons that can be explained using the combination of the number of water molecules in the hydrate phase as a function of time, $n_{h}=n_{h}(t)$, and the number of wells filled by water molecules as a function of time, $N_{fw}=N_{fw}(t)$. According to our previous experience on the determination of interfacial free energies of hydrates, we think the following plots could explain why we have chosen the $r_{w}^{0}$ values presented in the main manuscript.

\medskip

\begin{figure}
\centering
\includegraphics[width=0.35\textwidth]{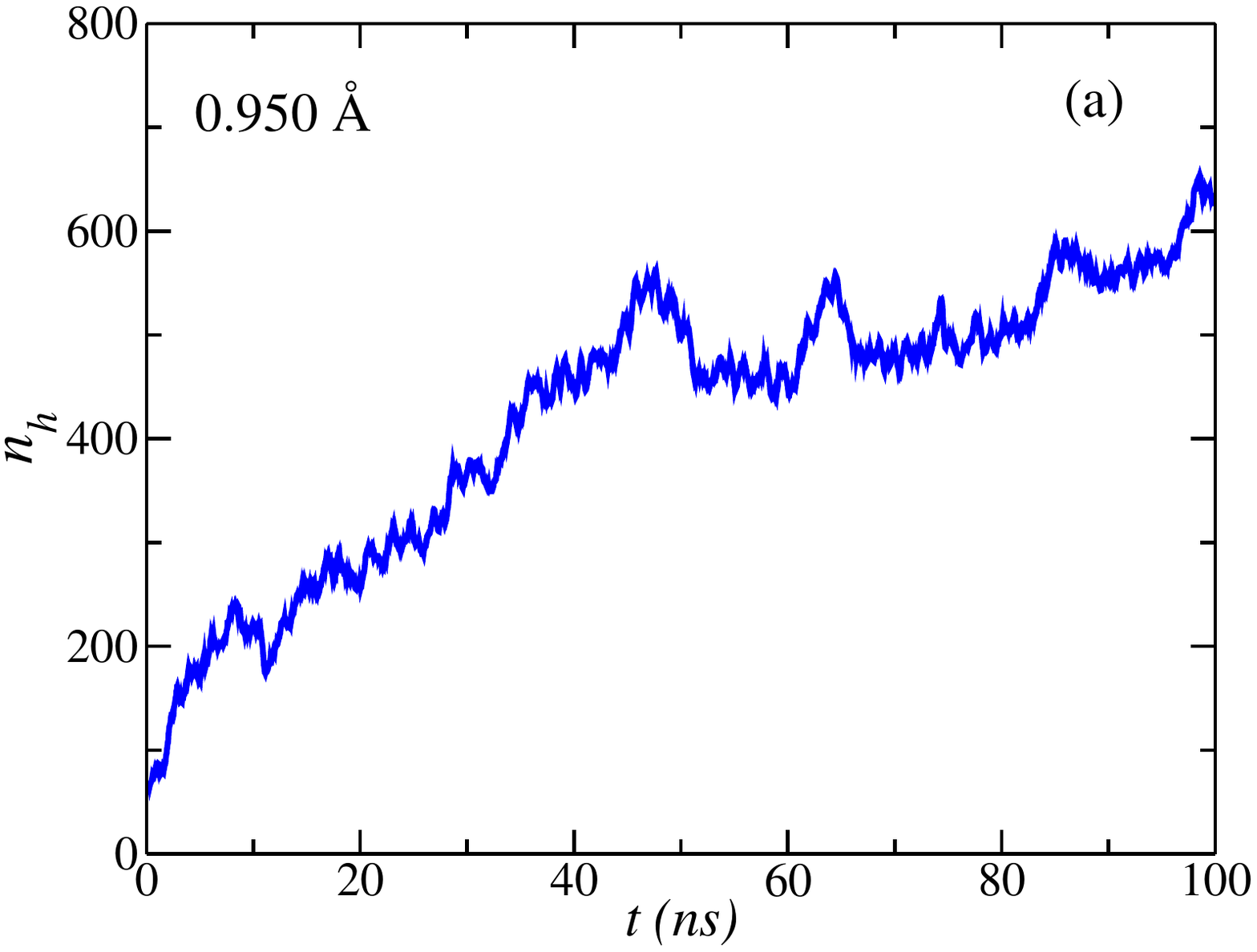}
\includegraphics[width=0.35\textwidth]{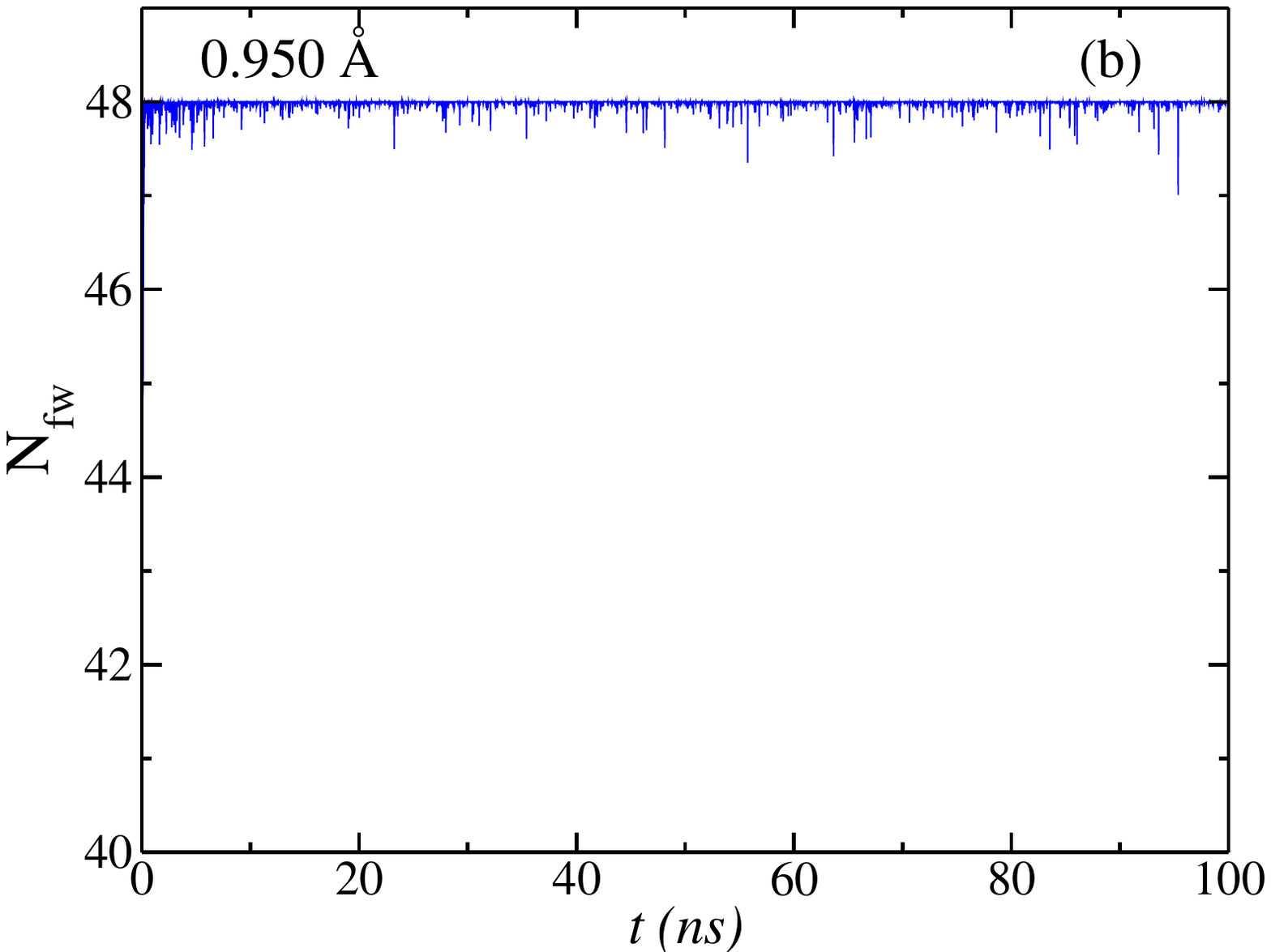}
\includegraphics[width=0.35\textwidth]{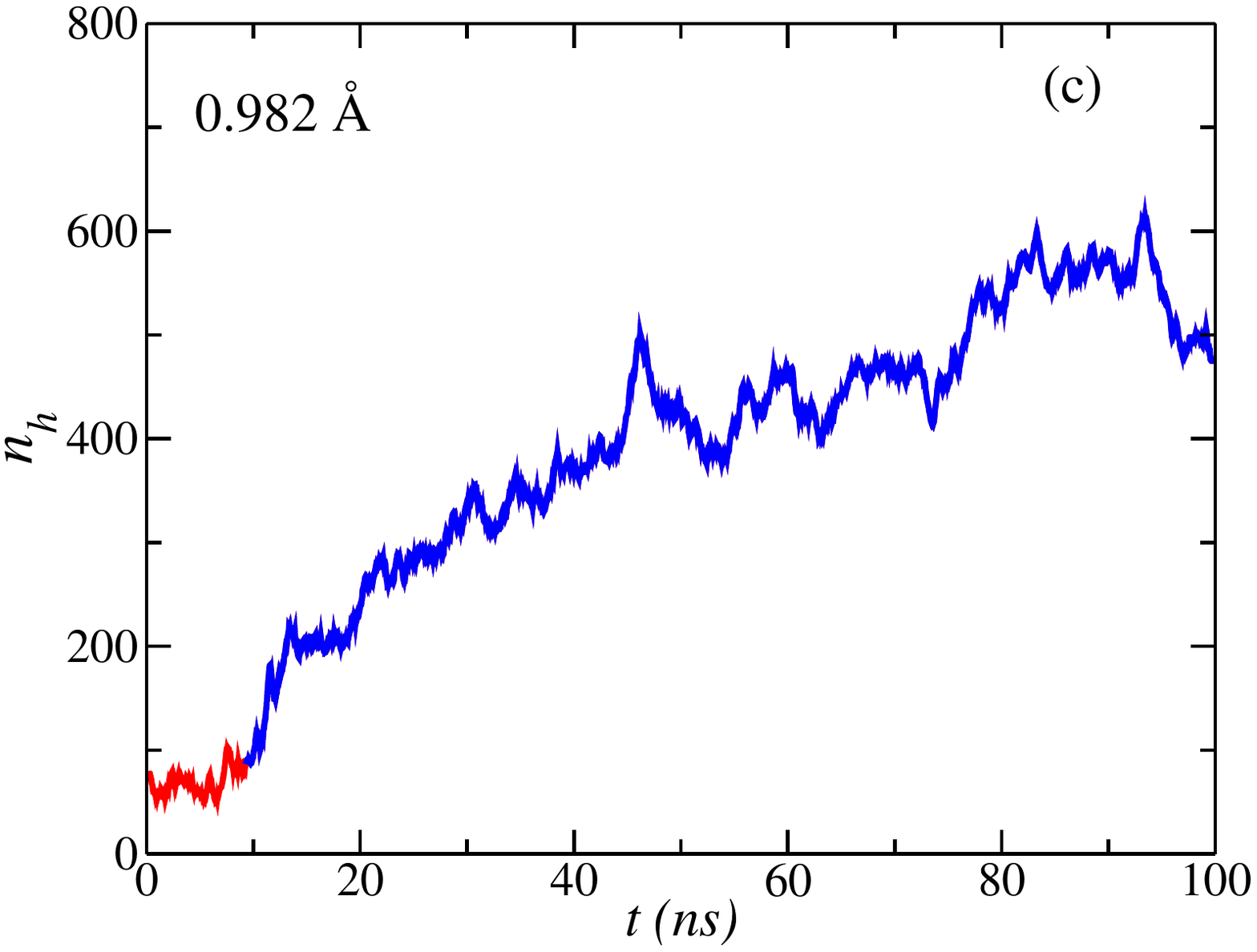}
\includegraphics[width=0.35\textwidth]{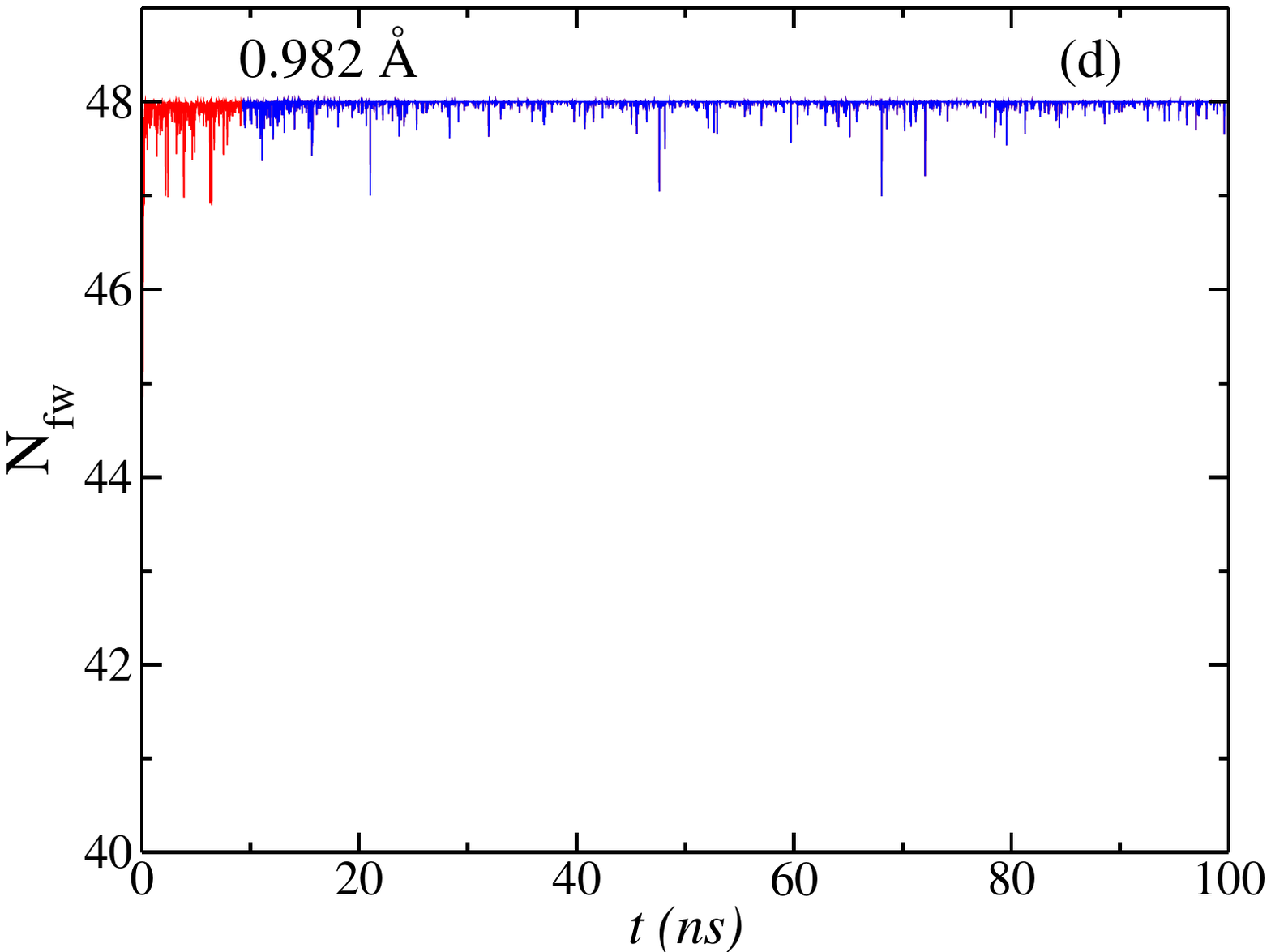}
\includegraphics[width=0.35\textwidth]{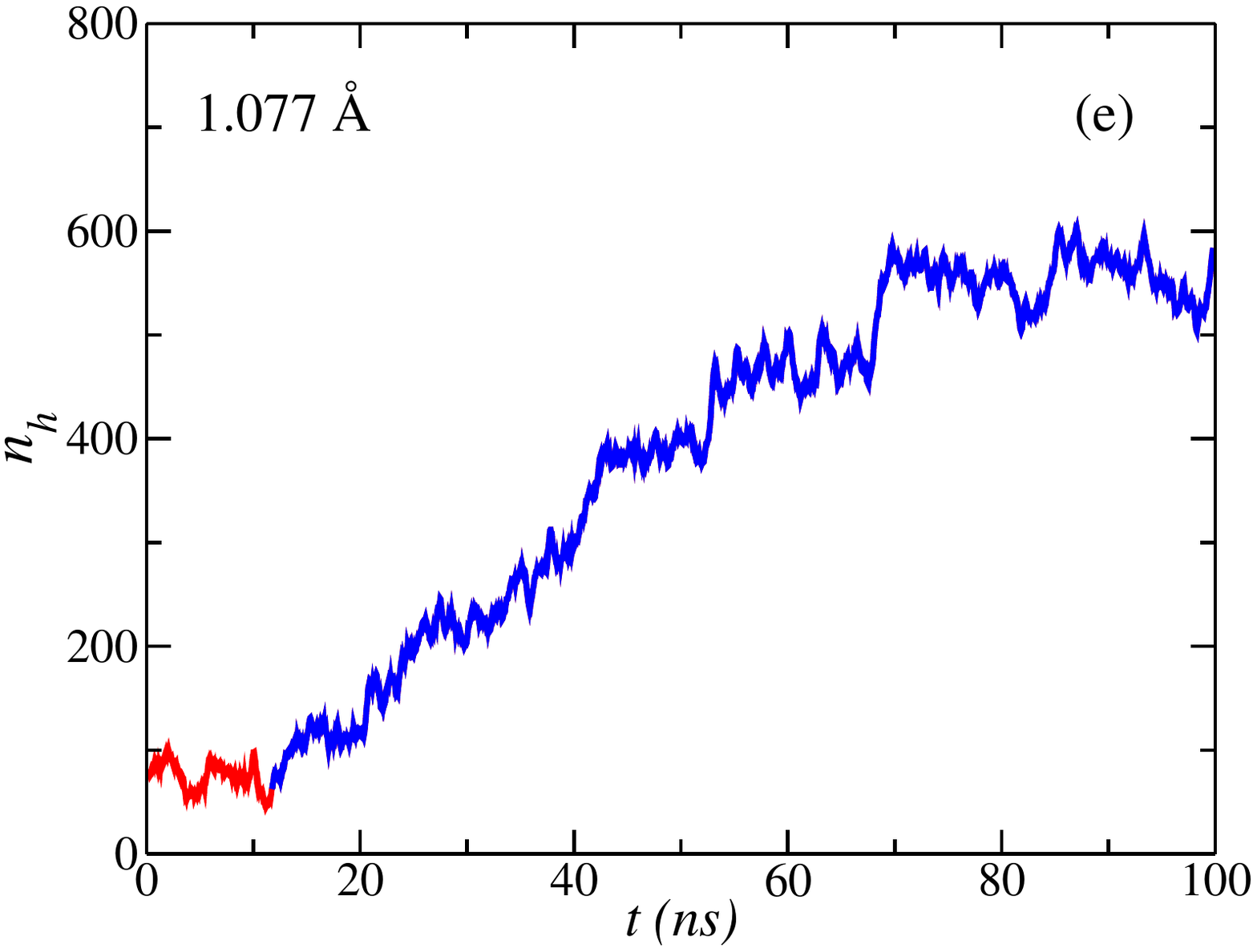}
\includegraphics[width=0.35\textwidth]{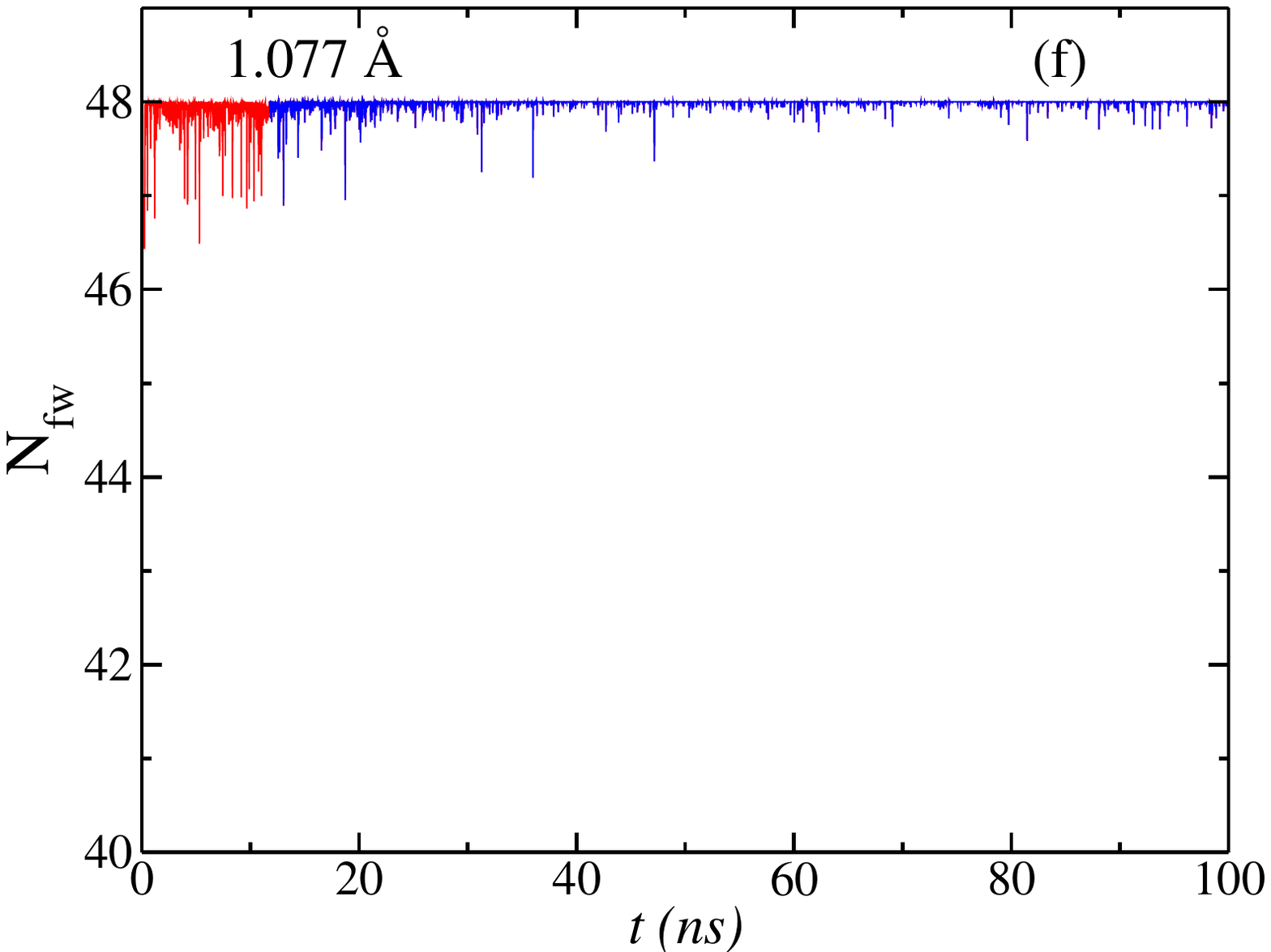}
\includegraphics[width=0.35\textwidth]{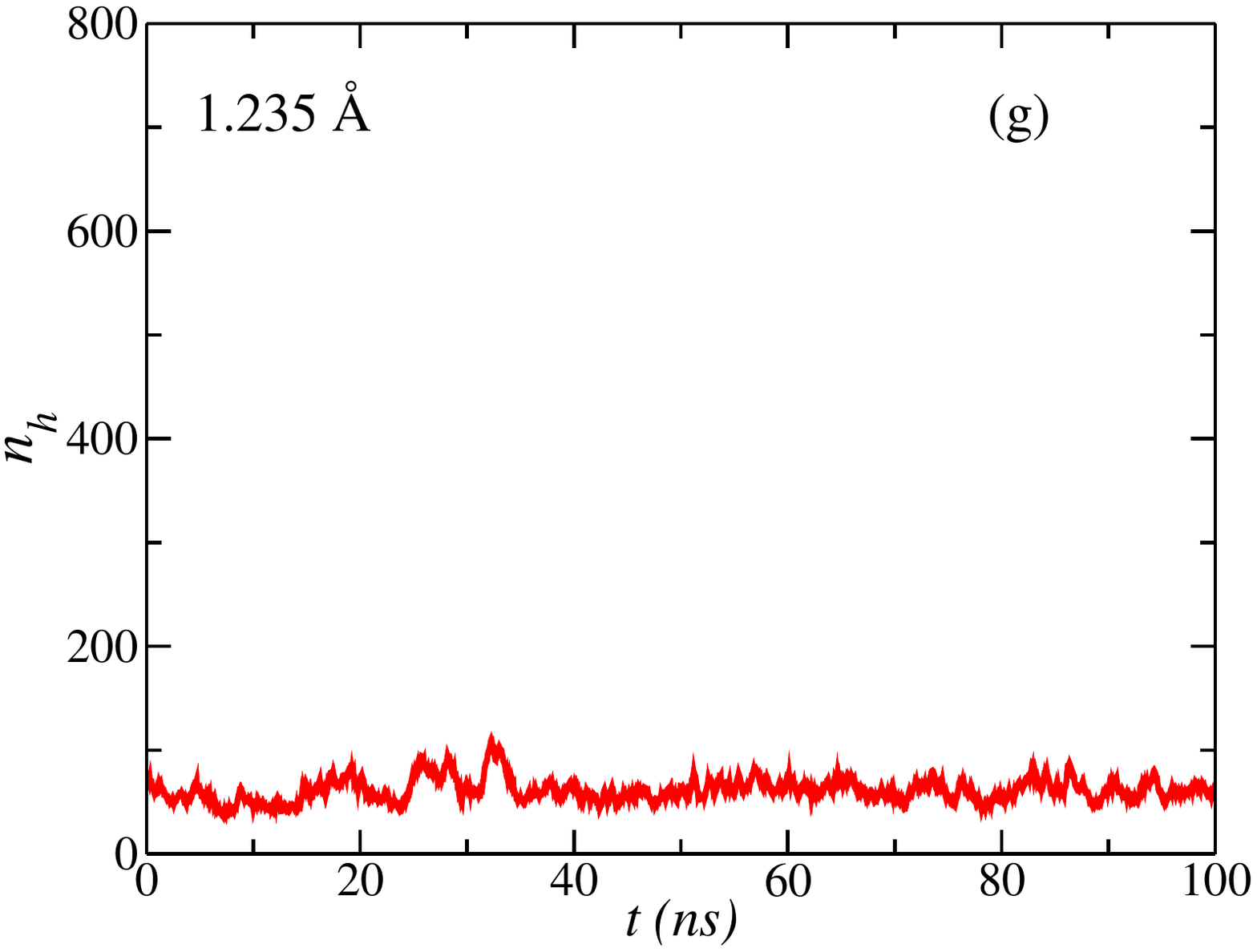}
\includegraphics[width=0.35\textwidth]{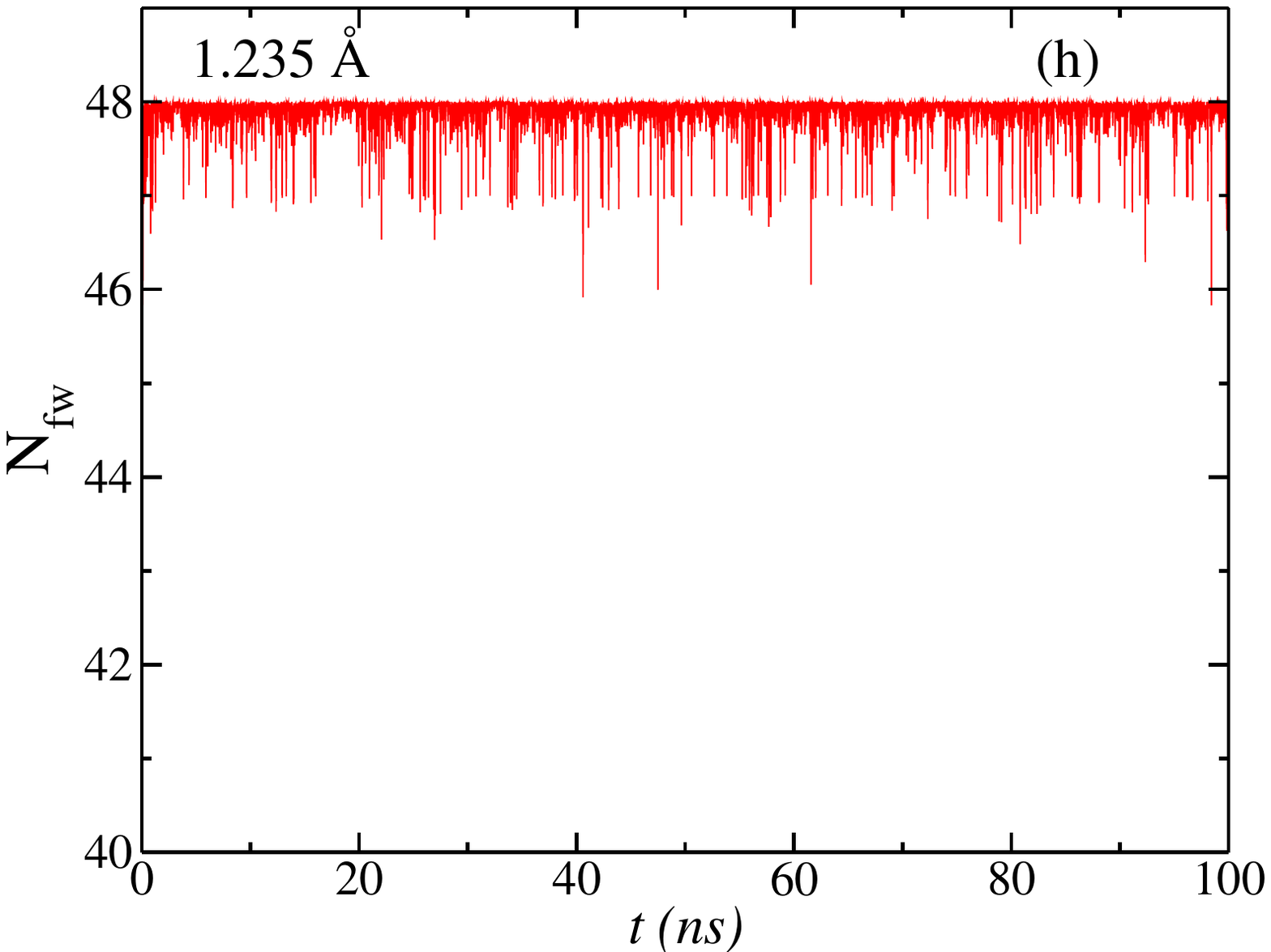}
\caption{\small Number of water molecules in the hydrate slab, $n_{h}$
as a function of time in panels (a), (c), (e), and (g); and number of filled wells, $N_{fw}$, as a function of time in panels (b), (d), (f), and (h), for some trajectories using different well radius $r_{w}$ (as indicated in the legends). Simulations are performed using $48$ wells, with $\varepsilon = 8\,k_{B}T$, and at coexistence conditions of $1000\,\text{bar}$ and $289\,\text{K}$. Blue colors represent trajectories or parts of trajectories classified without induction period (hydrate growing) and red colors are trajectories or parts of trajectories in which an induction period is identified.}
\label{figure1}
\end{figure}

Fig.~\ref{figure1} shows $n_{h}=n_{h}(t)$ and $N_{fw}=N_{fw}(t)$, for some trajectories at $1000\,\text{bar}$ already presented in the manuscript, that are relevant to the discussion about the existence of induction period. The discussion and results presented here are equivalent to those for $100$ and $400\,\text{bar}$. Particularly, we consider the following trajectories: trajectory without induction period using $r_{w}=0.950\,\text{\AA}$ ($<r_{w}^{0}$), panels (a) and (b); trajectory with induction period using $r_{w}=0.982\,\text{\AA}$ ($\sim r_{w}^{0}$), panels (c) and (d); trajectory with induction period using  $r_{w}=1.077\,\text{\AA}$ ($\sim r_{w}^{0}$), panels (e) and (f); and trajectory with an ``infinite'' induction period, i.e., trajectory that does not show crystallization during the complete simulation ($100\,\text{ns}$) using $r_{w}=1.235\,\text{\AA}$ ($>r_{w}^{0}$), panels (g) and (h). The key point of this figure is to show that the elections suggested by us make sense and are in agreement with the spirit of the original methodology. As can be clearly seen in panel (a), there is no induction period. Panel (b) shows the instantaneous number of wells filled by water molecules. The behavior of $N_{fw}$ as a function of time is the same (within the statistical uncertainties) in the whole $100\,\text{ns}$ of the simulation. Particularly, all the wells are filled ($48$) and only small fluctuations are seen along the complete simulation. The mean value of $\overline{N}_{fw}\sim 47.99$ and the standard deviation is $\sigma_{\overline{N}_{fw}}\sim 0.046$. It is important to mention that all the wells of the mold are filled in the first $200\,\text{ps}$ of the simulations in all the cases considered in this work.

\medskip

Panel (g) shows the opposite behavior: the system does not crystallize during the whole simulation ($100\,\text{ns}$), i.e., one can say that the system exhibits an ``infinite'' induction period. In other words, the free energy barrier between the hydrate phase and the aqueous solution is so high that the probability associated to a transition from the fluid phase to the hydrate phase is negligible. This behavior is reflected in panel (h) where $N_{fw}(t)$ shows a similar trend than in panel (b) but with an important difference: fluctuations of $N_{fw}$, $\sigma_{N_{fw}}$, are larger. According to our calculations, in this case $\sigma_{\overline{N}_{fw}}\sim 0.21$ (note that here $\overline{N}_{fw}\sim 47.88$). This happens when $r_{w}=1.235\,\text{\AA}$. Why fluctuations are $4-5$ times greater than in panel (b)? Only water molecules trapped in the wells of the mold form the thin slab of hydrate and no hydrate-like water molecules exist in the system (the system does not crystallize due to the height of the free energy barrier between the hydrate and the liquid). This provokes that a few percentage of molecules go out from some wells from time to time, increasing $\sigma_{N_{fw}}$. This does not occur in panel (b), for $r_{w}=0.950\,\text{\AA}$, because the system crystallizes and the hydrate phase grows around the region at which the wells of the mold are located in the simulation box. In the same way, ``the hydrate-like water molecules provide extra stability to the water molecules at the wells'' reducing the $\sigma_{N_{fw}}$ value.

\medskip

Panels (c) and (e) show qualitatively the same behavior: induction periods in the first $10\,\text{ns}$ of the simulation (red color) and hydrate growing (crystallization) in the rest of the simulation time (blue color). This is also corroborated by the values of fluctuations of $N_{fw}$ in the first $10\,\text{ns}$: $\sigma_{\overline{N}_{fw}}\sim 0.15$ in panel (d) and $\sigma_{\overline{N}_{fw}}\sim 0.19$ in panel (f) (trajectories in red color). According to panel (h), these values correspond to induction periods. And also in the rest of the simulation time:  $\sigma_{\overline{N}_{fw}}\sim 0.05$ in panel (d) and $\sigma_{\overline{N}_{fw}}\sim 0.05$ in panel (f) (trajectories in blue color). According to panel (c), these values correspond to crystallization periods. Consequently, we do not see real, qualitative, and quantitative differences between the behavior of $n_{h}(t)$ and $N_{fw}(t)$ in panels (c)+(d) and (e)+(f), respectively. Then, we assume that $r_{w}=0.982\,\text{\AA}$ is the lowest value of $r_{w}$ for which at least one trajectory exhibits induction period. Since $r_{w}=0.950\,\text{\AA}$ is the highest value for which none of the trajectories shows an induction period, $r_{w}^{0}\approx 0.966\,\text{\AA}$. This criterion is strictly the same one proposed by the authors in the original MI methodology~\cite{Espinosa2014a,Espinosa2016a} and the same one we have used in this work for $100$ and $400\,\text{bar}$ and in our previous works for determining the interfacial free energy of the CO$_{2}$ hydrate~\cite{Algaba2022b,Zeron2022a}. In order to keep the consistency with our previous works, we opt to follow the same approach for choosing the $r_{w}^{0}$ values.

\medskip
It is also important to remark that we have also found trajectories in which $N_{fw}(t)$ behaves during a certain time like in the red regions of panels (d), (f), and (h) but the system is crystallizing, i.e., $n_{h}(t)$ behaves during that time like in the blue regions of panels (c), (e), and (g). These results correspond to few simulations currently in course. Taking into account only the results presented in this work, our main conclusion is that information from the behavior of $n_{h}(t)$ alone or from the behavior of $N_{fw}(t)$ alone is not enough to identify if a trajectory has induction period or not. We think that a careful inspection of both magnitudes, $n_{h}(t)$ and $N_{fw}(t)$, helps to identify in most of cases if one trajectory exhibits or not induction period. Clearly, we think that a more detailed study of this issue is necessary. This is devoted to a future work since this is out of the scope of this paper.

\subsection{Overall behavior criterion}

According to the original criterion, in combination with the careful inspection of the number of filled wells and their fluctuations, the region with large fluctuations in the mold occupancy is identified with an induction period, and consequently, with the absence of a crystal structure around the mold. According to the original methodology, this implies that the system is in an induction period prior to the overcoming of a free energy barrier. However, another interpretation is possible. The absence of crystal structure could be also interpreted as a stuck crystal growth due to the low step-like kinetics with which the hydrate apparently grows. Unfortunately, it is very difficult to distinguish between induction period and stuck growth at the beginning of the simulations in which there is absence of crystal structure.

\medskip

\begin{figure}[t]
\centering
\includegraphics[angle=0,width=0.65\textwidth]{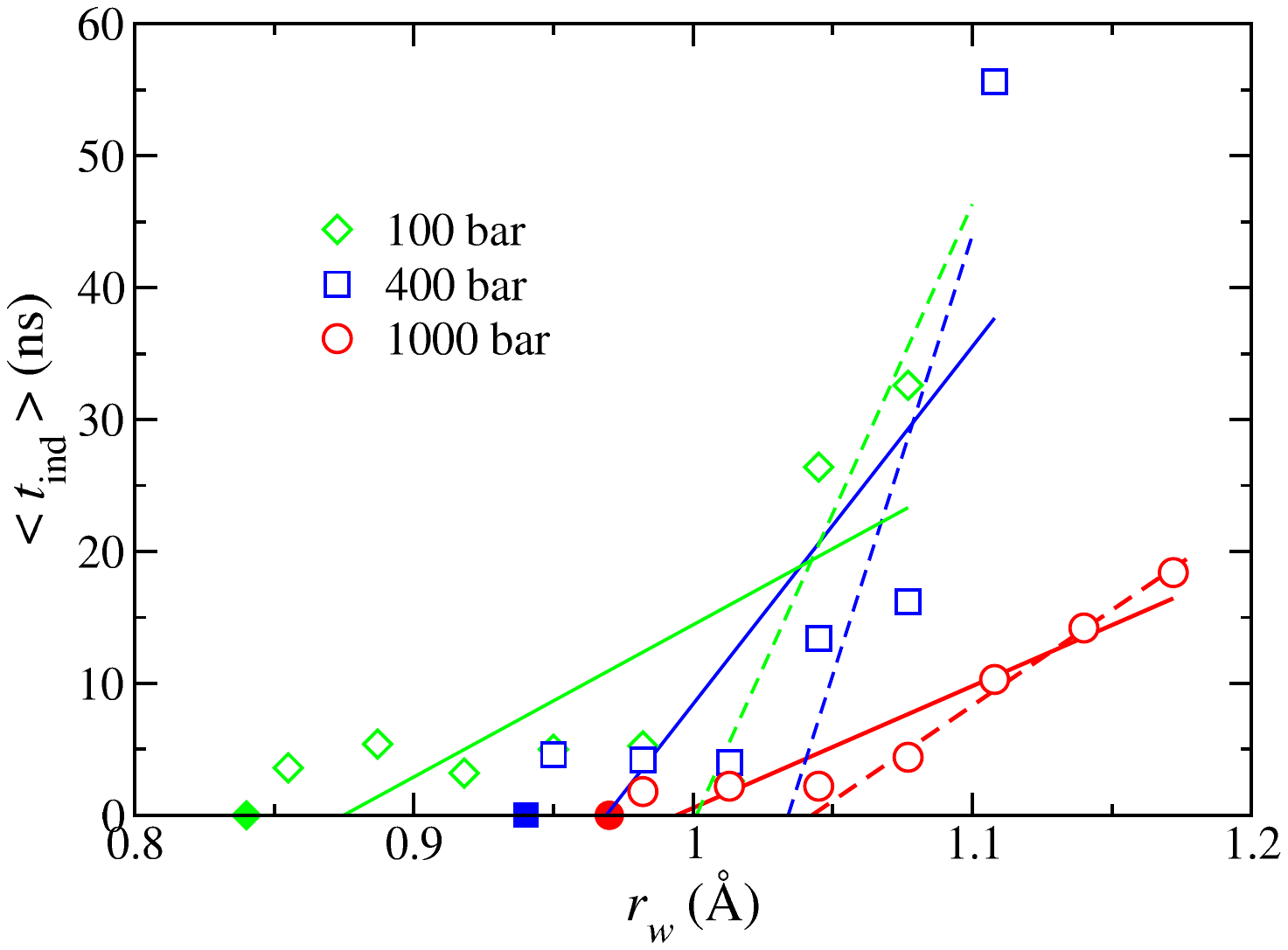}
\caption{\label{s8}\small Average induction period, $<t_{\text{ind}}>$, as a function of $r_{w}$ for pressures $100$ (green open diamonds), $400$ (blue open squares), and $1000\,\text{bar}$ (red open circle). The filled symbols correspond to the $r_{w}^{0}$ values obtained from the original criterion at $100$, $400$, and $1000\,\text{bar}$.}
\end{figure}

An option could be look at the overall behavior as a function of the well radius to discriminate between both views. To this end, Fig.~\ref{s8} shows the average induction period, $<t_{\text{ind}}>$, as a function of $r_{w}$, identified in Figs.~\ref{s1}-\ref{s6} for all the simulations considered in this work that exhibit possible induction periods higher than $3\,\text{ns}$. Data with different colors correspond to the three pressures considered in this work, $100$ (green), $400$ (blue), and $1000\,\text{bar}$ (red). The filled symbols along the $x$-axis represent the optimal radii, $r_{w}^{0}$, at each pressure obtained using the original criterion (see also Table~\ref{table}).

\medskip

The general behavior of each series is qualitatively the same. Each pressure shows two different regions: (1) a region (I) in which 
$<t_{\text{ind}}>$ is approximately constant, with $<t_{\text{ind}}>\sim 5-7\,\text{ns}$, for low values of $r_{w}$; and (2) a second region (II) in which $<t_{\text{ind}}>$ grows (approximately) linearly for high values of $r_{w}$. According to the definition of induction period, at each pressure, $r_{w}^{0}$ can be obtained from the overall behavior criterion extrapolating the induction period data towards zero, i.e., considering the limit $<t_{\text{ind}}>\rightarrow 0$. However, as we have previously mentioned, two different behaviors are seen in Fig.~\ref{s8}. One could be tempted to discard the $<t_{\text{ind}}>$ values of region I since these points correspond to states in which induction period and stuck growth can be confused, as we have previously mentioned. Note that these values correspond to simulations in which a possible induction period is observed according to the analysis of $n_{h}=n_{h}(t)$ (see Figs.~\ref{s1}-\ref{s6}). The behavior of the system at these states could be considered isolated events. However, this happens systematically. As can be seen in Fig.~\ref{s8}, these states exist at the three pressures and for several values of $r_{w}$.

\begin{figure}[t]
\centering
\includegraphics[angle=0,width=0.65\textwidth]{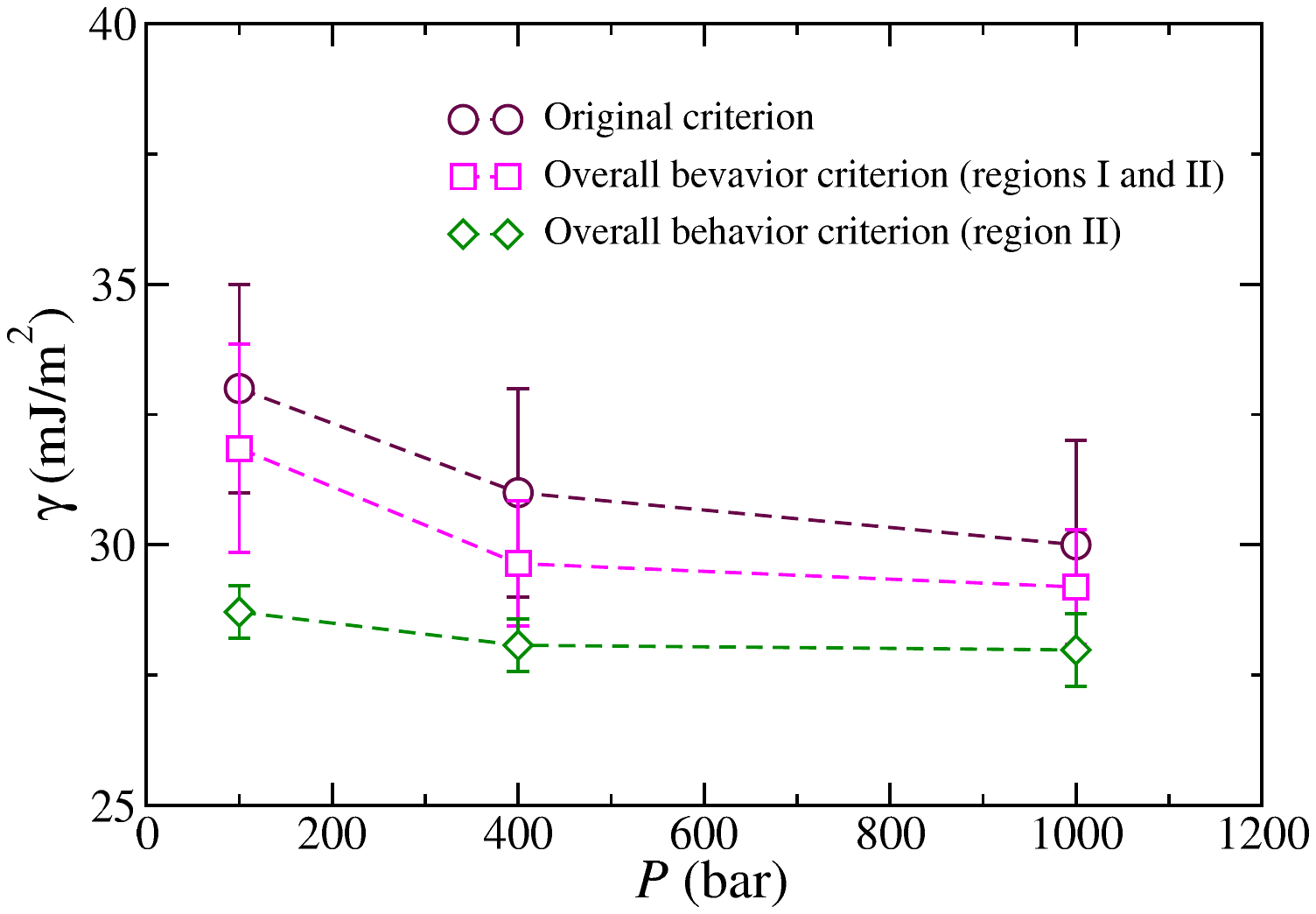}
\caption{\label{s9} \small CO$_{2}$ hydrate--water interfacial free energy as a function of pressure along its dissociation line as obtained from the MI--H methodology using the original criterion (marron symbols), the overall behavior criterion with all the data (magenta symbols), and the overall behavior criterion with only the data from region II (dark green). The dashed lines are guides to the eye.}
\end{figure}

To account for properly the existence of these states, we have performed two different linear fits from the data presented in Fig.~\ref{s8}. The first one considers all the $<t_{\text{ind}}>$ values (regions I and II), shown by continuous lines with different colors. We have also performed a second linear fit taking into account only the $<t_{\text{ind}}>$ values of region B, shown by dashed lines. The optimal values of $r_{w}$ obtained from both linear fits have been also included in Table~\ref{table}. We have also studied the corresponding uncertainties using the same criterion presented in the main article.

From this information, it is possible to calculate the interfacial free energy of the hydrate, at different pressures, using both criteria. We have represented the interfacial free energy as a function of pressure in Fig.~\ref{s9}. We have included the results obtained from the overall behavior criterion using the the two linear fittings (including only region I and including regions I and II). We have also plotted the results obtained using the original criterion. As can be seen, the three results show the same qualitative behavior: there is a weak correlation between the interfacial energy values and the pressure, i.e., $\gamma_{hw}$ slightly decreases with pressure along the dissociation line. These effect is more pronounced when using the original criterion and the overall behavior criterion with the linear fit including the data corresponding to region I and II. Finally, it is interesting to mention that results obtained from the original and overall behavior (including data from regions I and II) criteria predict the same results within the error bars.

\section{Average number of filled wells}

\begin{figure}
\centering
\includegraphics[angle=0,width=0.6\textwidth]{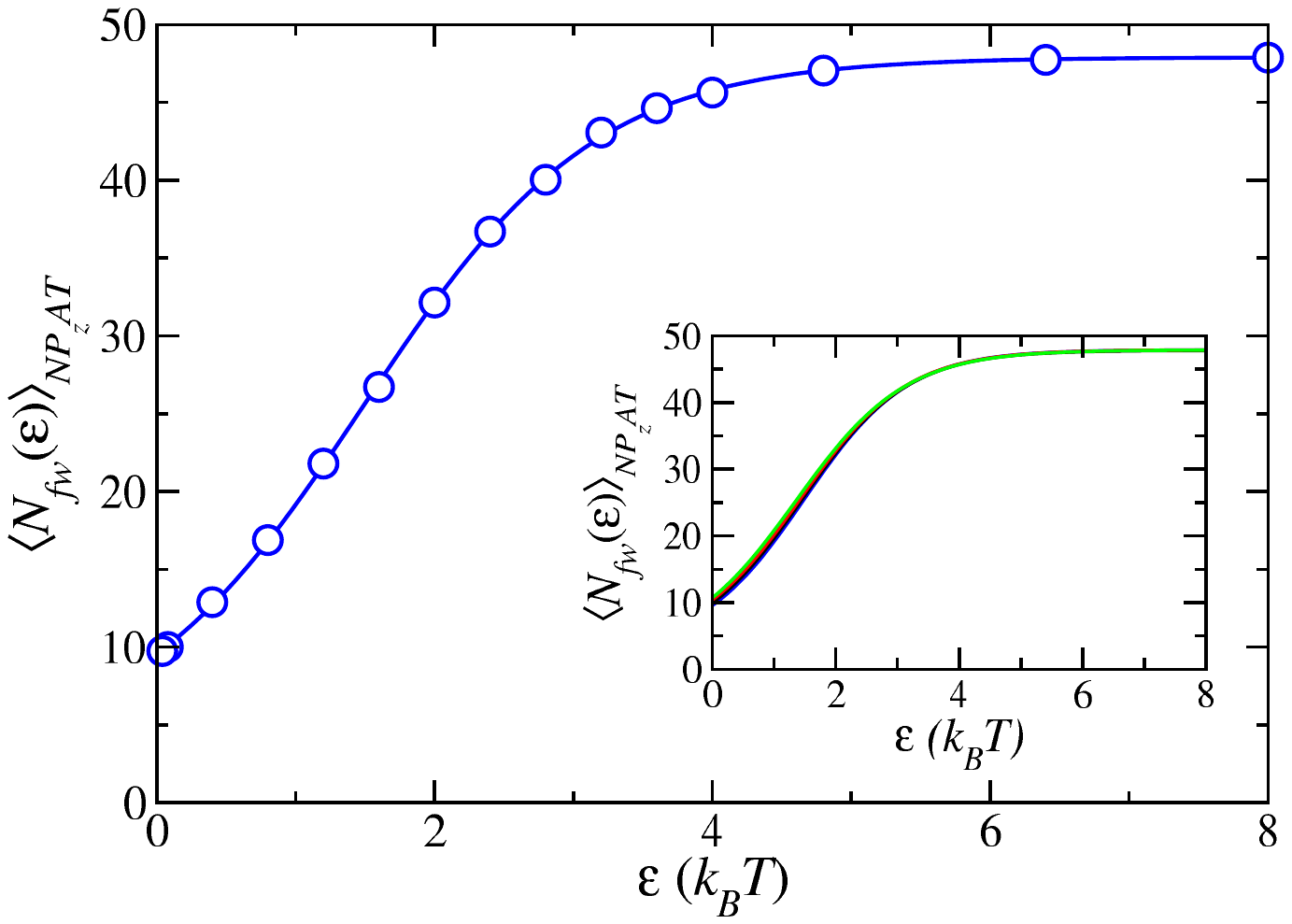}
\caption{\label{s10} Average number of filled wells, 
$\big \langle N_{fw}(\varepsilon)\big \rangle_{NP_{z}\mathcal{A}T}$, as a function of the well depth $\varepsilon$, at $400\,\text{bar}$ for the principal plane of the CO$_{2}$ hydrate. The radius of the mold used is $r_{w}=1.156\,\text{\AA}$. The circles correspond to the values obtained from $NP_{z}\mathcal{A}T$ simulations of $40\,\text{ns}$ and the blue curve represents its corresponding fit. The inset represents the fits of the average number of filled wells using well radii higher than the optimal value, $r_{w}=1.156$ (blue), $1.172$ (black), $1.188$ (red), and $1.203\,\text{\AA}$ (green).}
\end{figure}

\begin{figure}
\centering
\includegraphics[angle=0,width=0.6\textwidth]{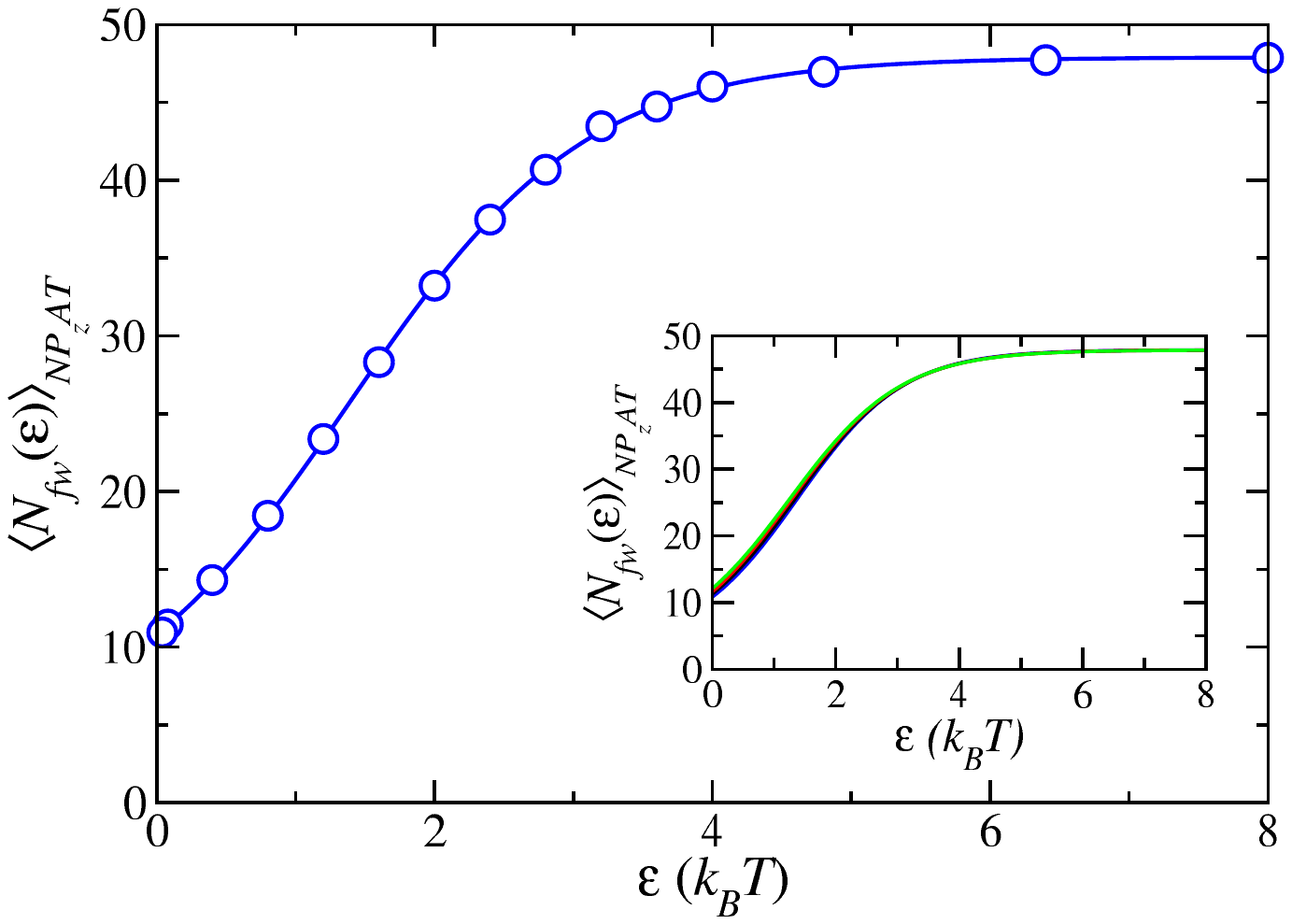}
\caption{\label{s11}Average number of filled wells, 
$\big \langle N_{fw}(\varepsilon)\big \rangle_{NP_{z}\mathcal{A}T}$, as a function of the well depth $\varepsilon$, at $1000\,\text{bar}$ for the principal plane of the CO$_{2}$ hydrate. The radius of the mold used is $r_{w}=1.203\,\text{\AA}$. The circles correspond to the values obtained from $NP_{z}\mathcal{A}T$ simulations of $40\,\text{ns}$ and the blue curve represents its corresponding fit. The inset represents the fits of the average number of filled wells using well radii higher than the optimal value, $r_{w}=1.203$ (blue), $1.219$ (black), $1.235$ (red), and $1.251\,\text{\AA}$ (green).}
\end{figure}

We have included the average number of filled wells, $\big \langle N_{fw}(\varepsilon) \big \rangle _{NP_{z}\mathcal{A}T}$, as a function of the well depth $\varepsilon$, at $400$ and $1000\,\text{bar}$ for the principal plane of the CO$_{2}$ hydrate. Fig.~\ref{s10} shows the filling curves, at $400\,\text{bar}$, using a value $r_{w}=1.156\,\text{\AA}$. We have also considered other values of the potential range of the wells, $r_{w}=1.172$, $1.188$, and $1.203\,\text{\AA}$, shown in the inset. Fig.~\ref{s11} shows the filling curves, at $1000\,\text{bar}$, using a value $r_{w}=1.203\,\text{\AA}$. We have also considered other values of the potential range of the wells, $r_{w}=1.219$, $1.235$, and $1.251\,\text{\AA}$, shown in the inset. Note that in all the cases $\big \langle N_{fw}(\varepsilon) \big \rangle _{NP_{z}\mathcal{A}T}$, as a function of time, behaves smoothly and it reaches the expected plateau value ($N_{fw}\rightarrow N_{w}$) when $\varepsilon\rightarrow\varepsilon_{m}$. As in the case of $100\,\text{bar}$ (main article), all the wells of the mold are fully occupied with only one water molecule for all the values of $\varepsilon$ used for the integration.

\clearpage

\clearpage

\section{Multimedia content}

We have included here the list of complementary multimedia results (movies) obtained from calculations performed in this work. In particular, these results correspond to Molecular Dynamics (MD) computer simulations in the $NP_{z}\mathcal{A}T$ ensemble starting from an initial and equilibrated liquid-liquid configuration of the CO$_{2}$ + H$_{2}$O binary mixture considered in the main paper. The movies shown here correspond to simulations carry out at $100\,\text{bar}$ and $284\,\text{K}$ using the Mold Integration - Host.

In our simulations, a mold of attractive sites is placed in the center of the simulation box, at the equilibrium positions of the oxygen atoms of water molecules along two principal planes of the sI structure of the CO$_{2}$ hydrate. We use
$N_{w}=48$ attractive interacting sites, with $\varepsilon=8\,k_{B}T$ and two different well values: $r_{w}=0.697\,\text{\AA}$, a value below the optimal radius value $r_{w}^{0}=0.84\,\text{\AA}$ (\textsf{video-MIH-crystallization.mpg}), and $r_{w}=1.172\,\text{\AA}$, a value
above the optimal radius value (\textsf{video-MIH-no-crystallization.mpg}). In the first case, the system crystallizes completely since there is no free energy barrier between the hydrate and the aqueous solution. In the second case, the system is not able to overcome the free energy barrier but is able to form a thin hydrate slab.

Red and white licorice representation corresponds to oxygen and hydrogen atoms of water, respectively. In the movies obtained using the MI-H methodology, blue and yellow spheres (van der Waals representation) correspond to carbon and oxygen atoms of CO$_{2}$, respectively. Finally, the 48 cyan spheres (van der Waals representation) correspond to the mold attractive sites using $\varepsilon=8\,k_{B}T$ and $r_{w}=0.697$ (crystallization) and $1.172\,\text{\AA}$ (no crystallization).

Here we provide the list of the files containing the movies indicating the time simulated in each case:

\begin{itemize}
\item Simulation using the MI-H methodology with crystallization ($100\,\text{ns}$):

\textsf{video-MIH-crystallization.mpg}.

\item Simulation using the MI-H methodology without crystallization ($100\,\text{ns}$):

\textsf{video-MIH-no-crystallization.mpg}.

\end{itemize}

\bibliography{bibfjblas}